\documentclass{emulateapj}

\newcommand{\zs}{$z$$\sim$}
\newcommand{\ze}{$z$$=$}
\newcommand{\zlya}{$z($Ly$\alpha)$}
\newcommand{\zeff}{$z_{\rm eff, Ly\alpha}$}
\newcommand{\lya}{Ly$\alpha$}
\newcommand{\Lya}{Ly$\alpha$}

\newcommand{\Ha}{H$\alpha$}

\newcommand{\sqam}{arcmin$^2$}

\newcommand{\sqdeg}{deg$^2$}
 
\newcommand{\cgsflux}{erg\,cm$^{-2}$\,s$^{-1}$}
\newcommand{\fluxcgs}{erg\,cm$^{-2}$\,s$^{-1}$}
\newcommand{\kmpersec}{km\,s$^{-1}$}
\newcommand{\phistar}{$\phi^*$}
\newcommand{\Lstar}{$L^*$}
\newcommand{\BRI}{$BRI$}
\newcommand{\Rab}{$R_{AB}$}

\newcommand{\nii}{[\ion{N}{2}]}
\newcommand{\sii}{[\ion{S}{2}]}

\shorttitle{Lyman $\alpha$ Emitters in DEEP2}
\shortauthors{Sawicki et al.}

\begin{document}


\title{The DEEP2 Redshift Survey: Lyman Alpha Emitters in the Spectroscopic Database\altaffilmark{1}}

\author{
Marcin Sawicki\altaffilmark{2,3}, 
Brian~C.~Lemaux\altaffilmark{4,5}, 
Puragra Guhathakurta\altaffilmark{4}, 
Evan~N.~Kirby\altaffilmark{4},
Nicholas~P.~Konidaris\altaffilmark{4},
Crystal~L.~Martin\altaffilmark{3,6},
Michael~C.~Cooper\altaffilmark{7},
David~C.~Koo\altaffilmark{4},
Jeffrey~A.~Newman\altaffilmark{8},
and 
Benjamin~J.~Weiner\altaffilmark{7}
}

\altaffiltext{1}{Based on data obtained at the W.M.\ Keck Observatory, which is operated as a scientific partnership among the California Institute of Technology, the University of California, and NASA, and was made possible by the generous financial support of the W.M.\ Keck Foundation.}

\altaffiltext{2}{Department of Astronomy and Physics, Saint Mary's University, 923 Robie Street, Halifax, Nova Scotia, B3H 3C3, Canada}

\altaffiltext{3}{Department of Physics, University of California, Santa Barbara, CA 93106, USA}

\altaffiltext{4}{University of California Observatories/Lick Observatory, Department of Astronomy and Astrophysics, University of California at Santa Cruz, 1156 High Street, Santa Cruz, CA 95064, USA}

\altaffiltext{5}{Department of Physics, University of California at Davis, One Shields Ave. Davis, CA 95616, USA}

\altaffiltext{6}{Packard Fellow}

\altaffiltext{7}{Steward Observatory, University of Arizona, 933 N.\ Cherry Av., Tucson AZ 85721, USA}

\altaffiltext{8}{Department of Physics and Astronomy, University of Pittsburgh, 3941 O'Hara St., Pittsburgh, PA, 15260, USA}

\slugcomment{ApJ, in press}

\begin{abstract}
We present the first results of a search for \lya\ emitters (LAEs) in the DEEP2 spectroscopic database that uses a search technique that is different from but complementary to traditional narrowband imaging surveys.  We have visually inspected $\sim$20\% of the available DEEP2 spectroscopic data and have found nine high-quality LAEs with clearly asymmetric line profiles and an additional ten objects of lower quality, some of which may also be LAEs.  Our survey is most sensitive to LAEs at $z$=4.4--4.9 and that is indeed where all but one of our high-quality objects are found. We find the number density of our spectroscopically-discovered LAEs to be consistent with those found in narrowband imaging searches. The combined, averaged spectrum of our nine high-quality objects is well fit by a two-component model, with a second, lower-amplitude component redshifted by $\sim$420~\kmpersec\ with respect to the primary \lya\ line, consistent with large-scale outflows from these objects.  We conclude by discussing the advantages and future prospects of blank-sky spectroscopic surveys for high-$z$ LAEs. 
\end{abstract}


\section{INTRODUCTION}

Over the last decade, studies of how galaxies form and evolve have been pushed to unprecedented look-back times, allowing us to observe galaxies at epochs when the universe was only a fraction of its present age.  A number of surveys for high-redshift ($z$$>$2) galaxies has been undertaken, including many in the rest-frame UV/optical (e.g., Steidel et al.\ 1996, 1999, 2003; Sawicki et al.\ 1997; Franx et al.\ 2003; Daddi et al.\ 2004; Sawicki \& Thompson 2005; Iwata et al.\ 2007) and far-IR (e.g., Barger et al.\ 1998; Hughes et al.\ 1998; Blain et al.\ 1999; Smail et al.\  2002; Chapman, et al.\ 2003; Webb et al.\ 2003; Sawicki \& Webb, 2005; Coppin et al.\ 2006). The study of galaxies selected by their strong \lya\ emission allows an investigation of a population that's potentially very different from these other high-$z$ galaxy populations.   \lya\ selection not only gives us the prospect of probing galaxies with very faint spectral continuum levels, but also may select systems in very early stages of a starburst (e.g., Malhotra \& Rhoads 2002).  It is thus not surprising that over the last several years much observational effort has gone into \lya\ surveys (e.g., Ajiki et al.\ 2003; Hu et al.\ 1998, 2004; Kudritzki et al.\ 2000; Ouchi et al.\ 2003;  Rhoads et al.\ 2000, 2003; Rhoads \& Malhotra 2001;  Santos et al.\ 2004). Indeed, the short rest-frame wavelength of the \lya1216 line allows galaxies to be detected and spectroscopically confirmed to very high redshifts, $z$$\lesssim$7, without the need for infrared observations.  

The ability to push to such high redshifts raises the intriguing possibility of probing the tail end of the epoch of reionization, regarded by many as marking the onset of galaxy formation in the Universe.  The expected evolution of the IGM neutral fraction across the reionization epoch has led to the prediction that the observed \lya\ luminosity function (LF) will undergo rapid evolution as IGM opacity to \lya\ photons drops with decreasing redshift (e.g., Haiman \& Spaans 1999; Malhotra \& Rhoads 2004; Kashihawa et al.\ 2006) and attempts have been made to detect the signature of this effect albeit with contradictory results (Malhotra \& Rhoads 2004; Kashikawa 2006).  At slightly lower redshifts, the LF of high-$z$ galaxies may be evolving due to evolution in the intrinsic properties of galaxies (e.g., Sawicki \& Thompson 2006; Iwata et al.\ 2007; see also Bouwens et al.\ 2007). Consequently even if the observed evolution of the \lya\ LF proves to be real, it will remain an open question as to whether it reflects an evolution in the opacity of the IGM, or evolution in the \emph{intrinsic} properties of galaxies at a time when they were young.  To believe the former we must understand the latter, and thus it is important to study the properties of Lyman Alpha Emitters (LAEs) in detail not only at very high redshift, $z$$>$6, but also as a function of time.

To date, narrowband imaging searches for high-$z$ LAEs have been the most fruitful approach to finding LAEs and have yielded significant samples of galaxies at \zs3--6.5, with several dozen objects now confirmed spectroscopically by various teams (e.g.,  Rhoads et al.\ 2003; Hu et al.\ 2004; Dawson et al.\ 2004, 2007; Shimasaku et al.\ 2006; Kashikawa et al.\ 2006).   An alternative to narrowband imaging surveys lies in direct, blank-sky spectroscopic searches whose strategy is to bypass the photometry stage and find galaxies directly in blank-sky spectra. One advantage of this approach is the potential saving of the time associated with deep narrowband imaging; another is the gain in sensitivity that results from working at the natural linewidth of the line ($\sim$10--20\,\AA) rather than fighting against the sky background admitted by the full width of the narrowband filter ($\sim$100--200\,\AA) (see Martin \& Sawicki 2004).   

Until recently, the chief disadvantage of spectroscopic searches has been the small sky area accessible to most spectrographs. Serendipitous spectroscopic discoveries of single emission-line objects (e.g., Franx et al.\ 1997; Dawson et al.\ 2002), while useful, do not yield {\it samples} of galaxies suitable for statistical analysis.  Spectroscopic searches of regions enhanced by gravitational lensing have met with some success (Santos et al.\ 2004; Stark et al.\ 2007), probe very deep into the luminosity function but over very small areas.  Until recently, multislit blank-sky searches (e.g., Crampton \& Lilly 1999; Martin \& Sawicki 2004; Tran et al.\ 2004) have found many line emitters but no confirmed LAEs;  only now are the sky areas accessible to this technique becoming large enough to report successful detections (Martin, Sawicki, Dressler, \& McCarthy 2008). 

In the present work we take an approach that is hybrid to the dedicated blank-sky spectroscopic surveys and the serendipitous discoveries.  Specifically, we carry out a dedicated search for serendipitous LAEs in the extremely large, existing spectroscopic database of the DEEP2 survey.  The present paper reports our first discovery of a sample of \lya\ emitters using these data and illustrates some of the constraints that this approach can place on the properties of these objects.  The paper is organized as follows.  In \S\ref{data.sec} we describe our DEEP2 data and their suitability to the task at hand.  In \S\ref{deep2serendips.sec} we describe our search for LAEs and report on the objects that we have found. In \S\ref{ConfidenceTests.sec} we present a number of tests with which we build confidence in the high-redshift nature of our objects. In \S\ref{numberdensity.sec} we constrain the number density and luminosity function of high-$z$ LAEs and in \S\ref{SpectralProperties.sec}  we study the spectral properties of the \lya\ emission line at high redshift. In \S\ref{discussion.sec} we discuss future prospects for finding LAEs using deep blank-field spectroscopy and summarize our main findings.  Throughout this paper we adopt the $\Omega_M$=0.3, $\Omega_\Lambda$=0.7, $H_0$=70~km~s$^{-1}$~Mpc$^{-1}$ cosmology.

\section{DATA}\label{data.sec}

Our study is based on an existing and very large database of two-dimensional spectra that were collected as part of the DEEP2 redshift survey.  We searched these spectra to identify emission-line objects that are not (other than by proximity in projection) associated with these primary DEEP2 spectroscopic targets. In this section we first describe the DEEP2 survey itself before launching into description of our search for serendipitous high-redshift galaxies in \S\ref{deep2serendips.sec}.

\subsection{The DEEP2 Survey}\label{deep2data.sec}

The DEEP2 survey (Davis et al.\ 2003, 2005; Coil et al.\ 2004) is a large spectroscopic redshift survey designed to measure redshifts and other properties of $\sim$50,000 galaxies using DEIMOS (DEep Imaging Multi-Object Spectrograph; Faber et al.\ 2003) on the 10m Keck II telescope. The primary goal of DEEP2 is the study of galaxy evolution at intermediate redshifts, \zs1.

The survey spans four fields, with Field 1 (the Extended Groth Strip) being a strip of 0.5~\sqdeg\ and Fields 2, 3, and 4 each spanning 1~\sqdeg, although in Fields 2 and 4 spectroscopic coverage was completed over only $\sim$0.67~\sqdeg.  Primary targets for spectroscopy are selected in deep \BRI\ images taken with the CFH12K camera on the Canada-France-Hawaii Telescope (for details, see Coil et al.\ 2004, Davis et al.\ 2005).  These images are used to select galaxies for DEIMOS spectroscopy down to a limiting magnitude of \Rab\ = 24.1.  Except in Field~1, a two-color cut is also applied to exclude galaxies with redshifts $z$$<$0.75.

Spectroscopy is obtained in multislit mode through custom-designed slitmasks. Slits are 1.0\arcsec\ wide and around 140 slits are cut per DEIMOS mask. Slit lengths range from 2 to 67 arcsec, with 90\% of slits being between 3.7 and 11.3 arcsec long and only 0.4\% of them shorter than 3 arcseconds.  Typical exposure times are 1 hour, although, when observing conditions are sub-optimal, this exposure time is usually increased to achieve a consistent survey depth across all masks. The primary intermediate-redshift target galaxy in each slit occupies only part of the slit length, and substantial area in each slit is exposed to blank sky in order to allow for accurate sky subtraction in the spectral reduction process.  In our work this ``background'' blank sky becomes the primary resource as it is there that we search for high-redshift emission-line galaxies.

The DEEP2 survey uses a 1200 line mm$^{-1}$ grating, which covers a spectral range of 6400--9000\,\AA\ at the moderately high resolution of R=5000.  The spectroscopic data were pipeline-reduced using an IDL pipeline developed at UC-Berkeley (Cooper et al.\ 2008, in preparation\footnote{For a description of the DEEP2 spectroscopic pipeline see \\ {\tt{http://astro.berkeley.edu/$\sim$cooper/deep/spec2d/primer.html}} \\ and {\tt{http://astro.berkeley.edu/$\sim$cooper/deep/spec1d/primer.html}}}).  For our emission-line search we use the data that have been taken through the reduction process that includes standard CCD pre-processing followed by two-dimensional sky subtraction.  The sky subtraction process (S. Burles and D. Schlegel 2008, in preparation) works exceptionally well and brings the data very close to the Poisson sky limit, with very few sky-subtraction artifacts remaining. It is these 2-dimensional sky-subtracted spectra that we use to search for high-$z$ line-emitting galaxies.

\subsection{DEEP2 as a high-z \lya\ survey}\label{data:deep2ashighz.sec}

While the chief goal of the DEEP2 survey is the study of the assembly of galaxies at intermediate redshifts, its extensive wavelength coverage, high spectral resolution, and large area make it an excellent dataset to search for galaxies at much higher redshifts.

The DEEP2 spectral range of $\sim$6400--9000\,\AA\ allows us to search for LAEs in the redshift range $z$$\sim$4.2--6.6. The large wavelength coverage of each spectrum also allows us to immediately reject most foreground interlopers through the detection of associated spectral lines at other wavelengths.  Moreover, the spectral resolution of the data is high enough to resolve line shapes and, in fact, in many cases our candidates exhibit the asymmetric shape expected of the \lya\ line.  Indeed, at our resolution of R=5000 we can easily resolve the [\ion{O}{2}]3727 doublet that can be a major source of contamination in \lya\ surveys (see, e.g., Martin \& Sawicki 2004; Martin, Sawicki, Dressler, \& McCarthy 2006, 2008); while the doublet may sometimes be smeared due to galaxy internal kinematics, our high spectral resolution allows us to reject many, if not all, of the contaminating [\ion{O}{2}]3727 emitters. Finally, the availability of $BRI$ imaging allows us a further check on our candidates since any bona-fide $z$$\gtrsim$4 emitters must also be $B$-band drop-outs.  In contrast to most LAE surveys, the DEEP2 spectroscopic data are already in hand and allow us to perform all these rejection tests without the need for additional observations.

The DEEP2 survey has accumulated a very large slit area exposed to the sky, which, combined  with its wide wavelength coverage, results in a very large LAE survey volume.  Each DEEP2 DEIMOS mask contains around 140 1\arcsec-wide slitlets, an arrangement that, after accounting for DEIMOS' inter-chip gaps ---gives the same sky coverage as a 15.4 arcminute-long 1\arcsec-wide longslit --- namely 0.257 \sqam\ per mask. The full DEEP2 survey consists of 385 masks, giving a sky coverage of 98.8 \sqam. The present study is based on 83 masks, which corresponds to 21.6\% of that 385-mask total.  Most of the spectra surveyed in the present work come from Field 4 of the DEEP2 survey. The area surveyed in the present work is 21.6~\sqam, which is somewhat larger than the areas surveyed by the recent custom-designed LAE multislit spectroscopic searches of Martin \& Sawicki (2004; 5.1~\sqam) and Tran et al.\ (2004; 17.6~\sqam), but substantially smaller than the $\sim$200~\sqam\ of the current state-of-the-art multislit search of Martin et al.\ (2008).  However, while the dedicated multislit surveys typically target only a narrow spectral range of 100--200\,\AA, our search is sensitive to objects over $\sim$6400--9000\,\AA.  Consequently, our search spans a disproportionately larger volume than is suggested by a simple comparison of survey areas. Our search --- which uses only $\sim$20\% of the available DEEP2 data --- surveys a volume of 6.9$\times$10$^4$~Mpc$^3$ and is thus already comparable to the 4.5$\times$10$^4$~Mpc$^3$ of the state-of-the-art dedicated Magellan multislit survey of Martin et al.\ (2008). It is much bigger than the volumes covered by the earlier dedicated multislit searches. 

The top panel of Figure~\ref{volume_vs_z.fig} shows the volume over which the DEEP2 survey is sensitive to LAEs.  This volume was calculated from a representative sampling of actual DEEP2 masks by multiplying the spectral coverage of each slit in these masks by that slit's length and width, and converting the result into volume as function of wavelength. The upper curve shows the volume for the entire (385-mask) DEEP2 survey and the lower curve that of the present search that uses only 83 of the DEEP2 masks.  The gray vertical lines show the locations of bright night-sky emission lines.

Several interesting features can be seen in Fig.~\ref{volume_vs_z.fig}(a), and here we briefly explain their origins. The steeply sloping ends of the wavelength range are a result of summing over the many thousands of spectra that make up the survey: while each individual DEEP2 spectrum has a well-defined minimum and maximum wavelength, these wavelengths vary from spectrum to spectrum subject to the slit's location on the mask and hence the spectrum's position with respect to the edges of the DEIMOS CCDs.  Taken as an average over the survey, the resulting survey volume has endpoints that are smoothly declining rather than abrupt as would be the case for a single spectrum.  Next, the smooth, shallow decline of the survey volume per redshift interval over the bulk of the survey range ($z$=4.6--6) is primarily a reflection of the redshift dependence of comoving volume per unit sky area in the adopted cosmology.  And finally, the shallow dip in coverage at $\sim$7800\,\AA\ is a consequence of the inter-chip gap between the two rows of DEIMOS CCDs. As with the spectral endpoints discussed earlier, this coverage gap is a sharply defined wavelength region for each DEEP2 spectrum, but taken as a sum over all the DEEP2 spectra this feature becomes a broad, shallow dip.

The vertical gray lines in Fig.~\ref{volume_vs_z.fig}(a) mark regions of increased noise due to night sky lines. In most surveys, these sky lines would make most of the wavelength range unusable for line emitter searches and this is why most LAE surveys target the few spectral regions that are relatively skyline-free (e.g., $\sim$8200\,\AA, $\sim$9200\,\AA). However, the DEEP2 spectral resolution is sufficiently high that in our case the skylines contaminate only a fraction of the total wavelength range and we are able to search for line emitters in the many inter-skyline regions throughout $\sim$6400--9000\,\AA.  Figure~\ref{volume_vs_z.fig}(b) shows the cumulative volume of the present, 83-mask survey (integrated from lower towards higher redshift), with the dashed line representing the volume as it would be in the absence of skylines and the solid curve showing the volume once we assume that the regions of high sky noise are rendered unusable by the higher sky noise. At the DEEP2 resolution,  substantial useful volume remains even in the sky-line rich regions. Having said that, however, we note that the region $\lesssim$7200\,\AA\ is particularly skyline-free, resulting in a concentration of useful survey volume at $z$$\sim$4.6. It is at these redshifts that most of our highest-quality LAE candidates are found.

Overall, the DEEP2 survey provides a very impressive and unique database for line emitter searches.  It covers a large volume, giving us the prospect of finding statistically significant samples of objects.  It has a large wavelength baseline and thereby offers the potential for evolutionary studies.  It has deep $BRI$ imaging that helps in the rejection of foreground interlopers.  It reaches faint spectroscopic flux limits of a few $\times10^{-18}$ \cgsflux, as we illustrate in \S~\ref{deep2serendips.sec}.  And the moderately high spectral resolution of its data not only provides us with large skyline-free regions to search in, but also gives us the ability to immediately ascertain the identity of our LAE candidates by inspecting their line profiles without the need for additional, follow-up spectroscopy. 

\section{LYMAN ALPHA EMITTERS IN DEEP2}\label{deep2serendips.sec}

\subsection{The search}\label{LyAsearch}

We searched the two-dimensional sky-subtracted DEEP2 spectra for serendipitous line-emitting objects.  The first steps of the search, performed by one of us (BL) were to visually identify an initial sample of serendipitous emission lines and then to define a preliminary catalog of those most likely to be LAEs on the basis of a number of interloper-rejection criteria.  Subsequently, this preliminary catalog was carefully re-examined by three of us (BL, MS, PG) to arrive at a final, robustly conservative LAE catalog.  Our guiding principle has been to reject objects liberally rather than to retain objects about whose nature there was some question. This philosophy results in an underestimate of the number of true LAEs but this is a deliberate policy on our part.  The steps of our classification procedure are explained in the following paragraphs.

The first and most time-consuming step ($\sim$400 person-hours) in the search was the visual inspection of over 10,000 two-dimensional spectra, representing $\sim$20\% of the DEEP2 dataset.  This search focused on identifying emission lines that were clearly not associated with the primary spectroscopic DEEP2 targets.  The emission lines of interest here were either lines that were offset spatially along the slit from the primary target, or lines that were superimposed on the spectrum of the primary target but clearly did not match any lines consistent with the redshift of the primary.

Once a serendipitous candidate had been identified as described above, a number of tests were applied to ascertain its identity. First, we rejected from our LAE catalog any candidate (other than those few that were spatially coincident with foreground objects) that had one or more additional emission lines at the same spatial location in the 2D spectrum:  for a true LAE galaxy the DEEP2 wavelength coverage would not include any prominent lines other than the \lya, and thus any lines additional to the discovery line brand the candidate as a low-$z$ object. The rejection procedure involved both the inspection of the entire length of the 2D spectrum and a targeted inspection of special regions in the 2D and 1D spectra.  These special  regions were regions where other emission lines would be expected to lie if our candidate were not \lya, but, rather,  [\ion{O}{3}] at 5007\,\AA, H$\beta$ at 4861\,\AA, H$\alpha$ at 6563\,\AA, or the [\ion{O}{2}] doublet at 3727\,\AA.  We also verified that the candidate is not an [\ion{O}{2}] emitter by checking for the double-peaked profile associated with this doublet line. In some cases, motions within galaxies can smear the doublet, but for all but very high electron densities, smeared galactic [\ion{O}{2}] can be expected to have a profile that is asymmetric in the opposite sense to that of \lya\ because of the relative intensities of the [\ion{O}{2}]3726,3729 doublet components (e.g., Osterbrock 1989); such inverted [\ion{O}{2}] ratios are found to be very rare observationally (Weiner et al.\ 2006). Thus, to check for smeared [\ion{O}{2}], we verified that our candidates do not have the [\ion{O}{2}]-like blue-tailed line profiles and found none.  Next, we also checked that neither the 2D nor 1D spectra had continuum blueward of the candidate line.  Once the candidate met all of these spectral tests, we next inspected the CFH12K images to search for plausible broadband counterparts to further confirm that the candidate had no strong continuum emission at wavelengths blueward of the candidate emission line.  This broadband imaging test does not absolutely rule out any chance that a given candidate is an interloper but instead achieves a more limited, though still important goal of verifying that no evidence exists that it is one.  Finally the three (or more) raw DEIMOS spectral frames were inspected at the locations of each of the candidates to make sure that the candidate emission line was not a cosmic ray artifact.

\subsection{The catalog}\label{LyAcatalog}

At the end of the search and weeding procedure described in \S~\ref{LyAsearch}, out of an initial sample of several hundred serendipitous line emitters, we retained 19 single-line objects.  Each of these nineteen was then assigned a confidence class, with class 3 objects being highest-quality, and class 1 objects being lowest; in \S~\ref{ConfidenceTests.sec} we do further tests on the objects of the different classes to gain further confidence in our highest-quality (class 3 and 2) objects.  

The confidence classes are defined as follows (see also Table~\ref{coad-subsets.tab} for a summary).  Class 3 and 2 objects are our highest quality LAE candidates --- both classes pass all the low-$z$ interloper tests described in \S~\ref{LyAsearch} and are characterized by a clearly asymmetric line profile typical of \lya. Class 3 objects are spatially well-separated from foreground galaxies (the primary DEEP2 spectroscopic targets or serendipitous foreground objects). Class 2 objects are close in projection to a foreground galaxy and so their fluxes could potentially  be attenuated by passage through the intervening galaxies' ISM and/or be boosted by galaxy-galaxy lensing. Thus, both class 3 and 2 objects are very likely to be high-$z$ LAEs, but class 2 objects are excluded from certain parts of our further analysis because of potential uncertainty in their true luminosities.  In contrast to class 2 and 3 objects, class 1 objects do not show clear asymmetric line profiles, but otherwise are single-line objects that pass all the interloper-rejection tests described in \S~\ref{LyAsearch}; they are the least secure candidates that we retain in our catalog, noting that they probably represent a mixture of LAEs and lower-$z$ interlopers (\S\ref{ConfidenceTests.sec}). 

The properties of the 19 objects are presented in Table~\ref{catalog.tab}, where we list their names (ID), observed wavelengths, redshifts (assuming they are \lya), positions (assuming location is at the centre of the 1\arcsec-wide slit), limits on fluxes and luminosities (at their \lya\ redshift), confidence classes, and the redshifts of the primary DEEP2 targets.  Object names are derived from a combination of DEEP2 mask number (the first four digits) and slit number (the digits that follow the period); the first digit of the name identifies the DEEP2 field in which the object is located.   Fluxes and luminosities listed in Table~\ref{catalog.tab} are lower limits because all objects are affected by slit losses (which cannot be known for individual objects in view of the degeneracy between position and wavelength), while, additionally, some objects were observed in non-photometric conditions.  The DEEP2 spectra were not fluxed and so we derived flux calibrations based on known instrumental throughputs, which we tested by comparing observed broadband magnitudes with spectral continua in a larger sample of DEEP2 primary targets. On the basis of these comparisons, we believe the LAE flux limits reported in Table~\ref{catalog.tab} to be accurate to $\sim$20\%. 

\begin{figure}
\plotone{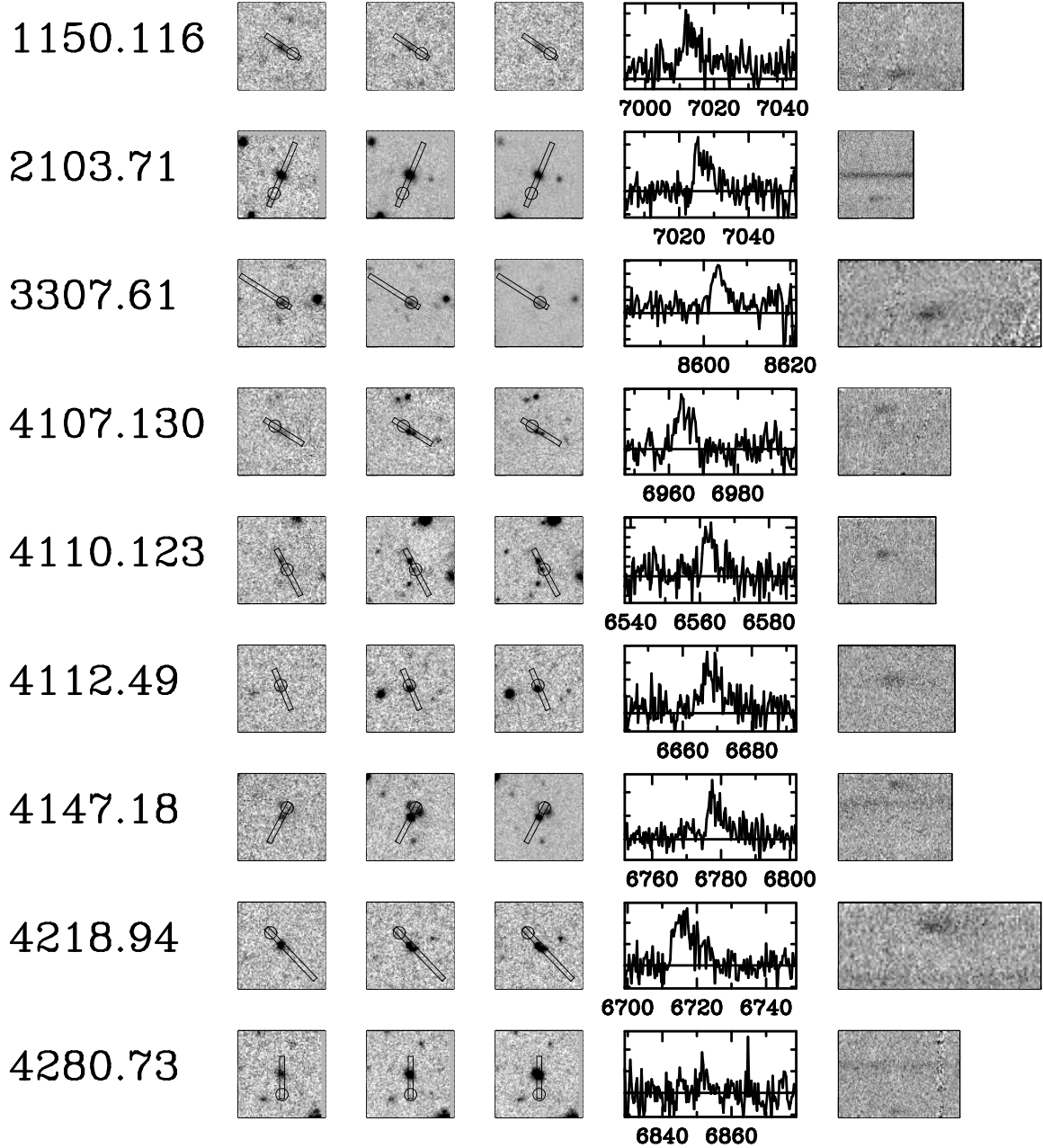}
\caption{\label{postagestamps.class23.fig} 
Direct images and spectra of our objects of confidence class 3 and 2.  The three left-most panels show direct $B$, $R$, and $I$ CFH12K images centered on the primary DEEP2 spectroscopic targets. Slit positions are indicated as are the inferred positions of our \lya\ candidates. These broadband images are 17.4\arcsec $\times$ 17.4\arcsec\ CFH12K image segments.  The right-most panels show the spectral images of the wavelength regions around the \lya\ candidates; wavelength increases to the right and the length of the image corresponds to 33\,\AA\ for each object, while the height of the image shows the full extent of the spectrum along the slit.  The 1-D spectra show the line shape at the position of the candidate line, with the wavelength scale in angstroms. }
\end{figure}

\begin{figure}
\plotone{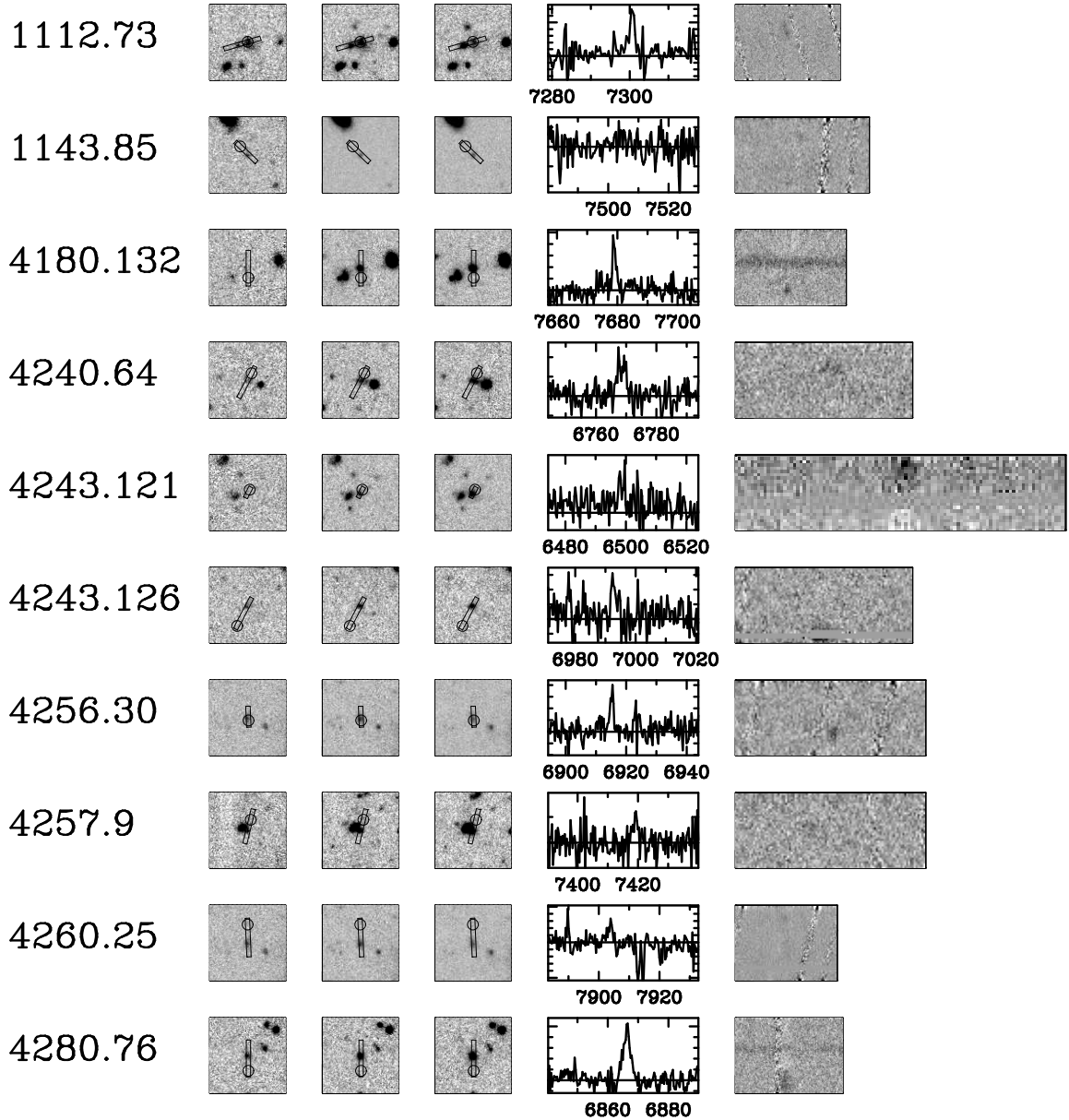}
\caption{\label{postagestamps.class1.fig} 
As for Fig.~\ref{postagestamps.class23.fig} but for the less secure objects of confidence class 1. }
\end{figure}

In Figs.~\ref{postagestamps.class23.fig}  and \ref{postagestamps.class1.fig} we show the direct images and spectra of our 19 objects.  The direct images show 17.4\arcsec $\times$ 17.4\arcsec regions and are taken from the $BRI$ CFH12K images. These images are centered on the positions of the primary DEEP2 spectroscopic targets, while positions of our line emitters are inferred from the spectroscopic data and are marked with open circles. The spectra in Figs.~\ref{postagestamps.class23.fig} and \ref{postagestamps.class1.fig} show both 2D spectral images and 1D spectral extractions of the region around the LAE.

\begin{figure}
\plotone{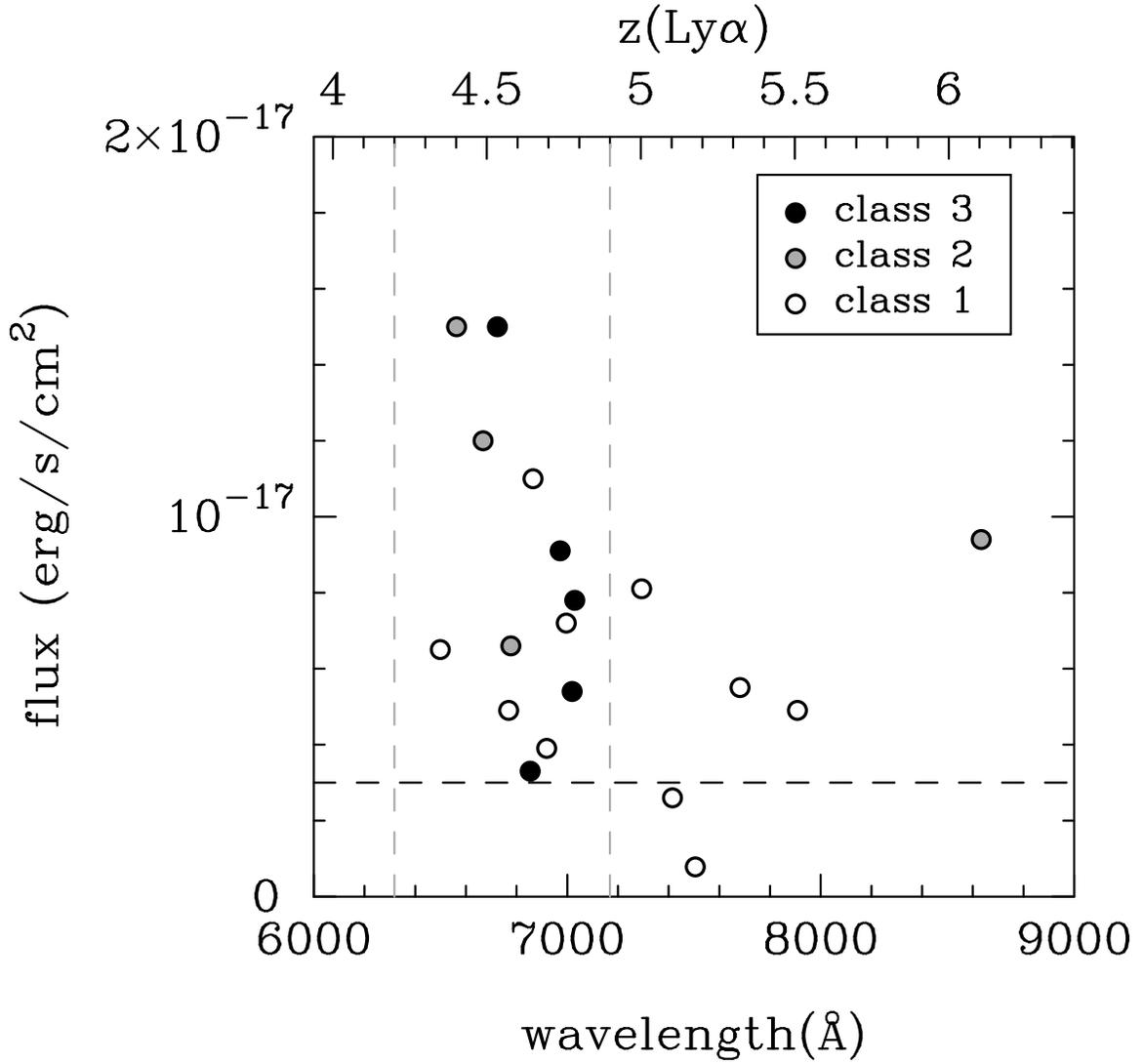}
\caption{\label{flux-vs-wav.fig} 
Observed fluxes and wavelengths of our line emitters.  The fluxes shown are lower limits on the true fluxes as they have not been corrected for slit losses.  The horizontal dashed line denotes our adopted limiting flux, and the vertical lines bracket the redshift range used in our number density calculations (\S~\ref{numberdensity.sec}).  Filled symbols show objects with the two highest \lya\ confidence classes.  }\end{figure}

Figure~\ref{flux-vs-wav.fig} shows the observed fluxes (i.e., lower limits on true fluxes) and redshifts of our LAEs.  Observed fluxes range from $\sim$5$\times$10$^{-17}$ \fluxcgs\ down to a few $\times 10 ^{-18}$ \fluxcgs.  Redshifts range over \zlya$\sim$4.3--6, as one would expect from the wavelength sensitivity of the survey as shown in Fig.~\ref{volume_vs_z.fig}.  Note that there is a concentration of class 3 and 2 objects at \zlya=4.4--4.8.  This concentration could be a result of real large-scale structure, but it could also reflect the larger skyline-free volume at these lower redshifts (Fig.~\ref{volume_vs_z.fig}), combined with the fainter luminosities accessible there as compared with higher redshifts. Since the bulk of our search was carried out in only one of the four DEEP2 fields, we will defer the analysis of clustering until the full DEEP2 dataset has been searched.

\section{Confidence Tests}\label{ConfidenceTests.sec}

The objects in our catalog of 19 galaxies have already passed all the individual tests described in \S~\ref{LyAsearch}.  In this section we present a number of tests that further build confidence in the high-$z$ nature of our sample.

\subsection{Averaged spectra}

\begin{figure}
\plotone{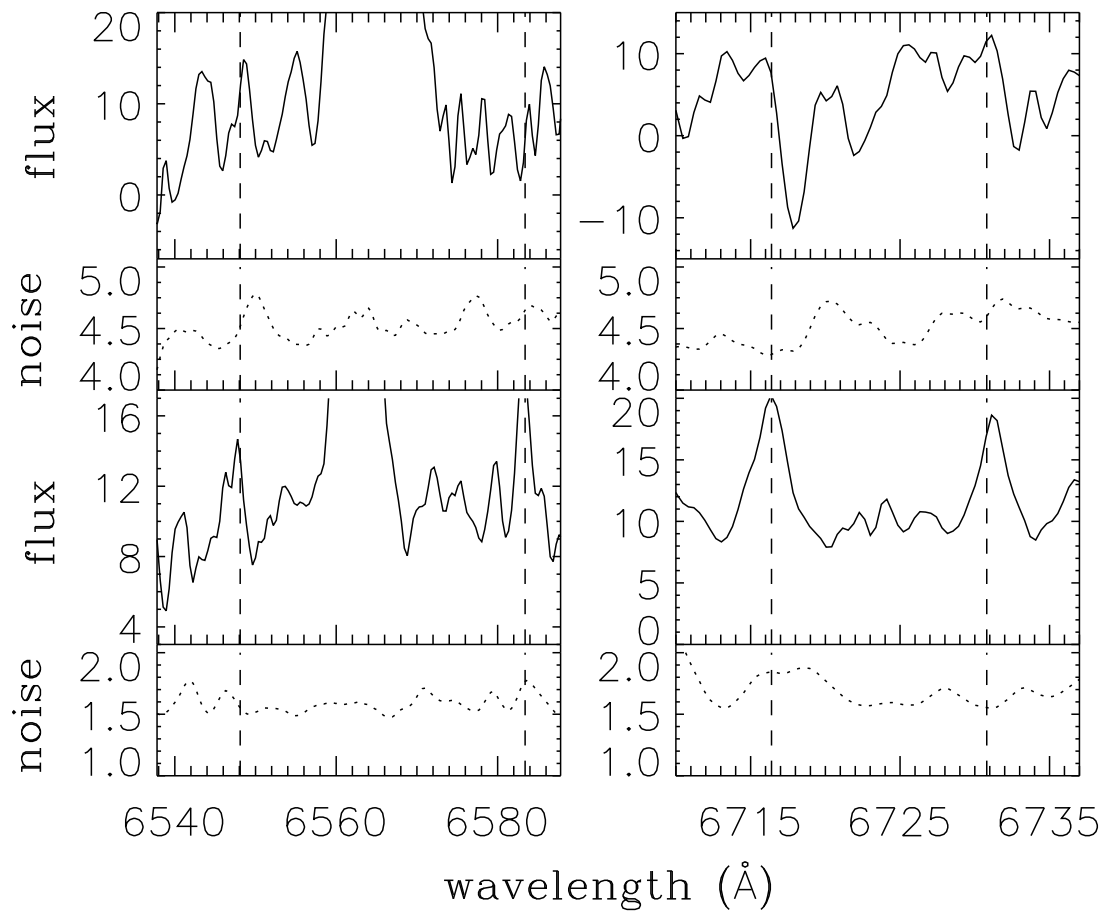}
\caption{\label{extralines.fig}
Spectral details for the co-added groups of spectra.   The top panels are for the composite of the 9 best candidates (class 3 and 2 objects), while the bottom panels are for the 61 non-LAEs single-line objects.  We assumed that we have detected \Ha\ and not \Lya\ and then use dashed lines to show the expected positions of [N~II] (left panels) and [S~II] (right panels). }
\end{figure}

Averaged spectra give higher signal-to-noise ratios than individual ones and thus can give additional confidence that, as a group, our candidates are genuine LAEs.  We constructed averaged spectra of three groups of objects: (a) Class 3 and 2 objects --- i.e., our best LAE candidates, (b) class 1 objects, and (c) 61 serendipitously-discovered DEEP2 single-line objects that have earlier been ruled out as LAEs through broad-band photometry (see Kirby et al.\ 2007).  These are shown in Figs.~\ref{lineshapes.fig} and~\ref{extralines.fig}.

To produce averaged spectra of each group, we shifted the peak of the line in each individual spectrum to a common rest-frame wavelength and then averaged each group of spectra with inverse variance weighting. We employed two types of averaging: uniform weighting and line weighting.  In the latter case, each spectrum is weighted by the signal-to-noise ratio of its emission line. Note that the class 1 object 1112.73 was not used in making the averaged spectra because this line emitter is too close to the principal DEEP2 target galaxy to allow for a reliable spectral extraction that's uncontaminated by flux from the foreground galaxy.

\subsubsection{Line shapes}\label{sec:lineshapes}

\begin{figure}
\plotone{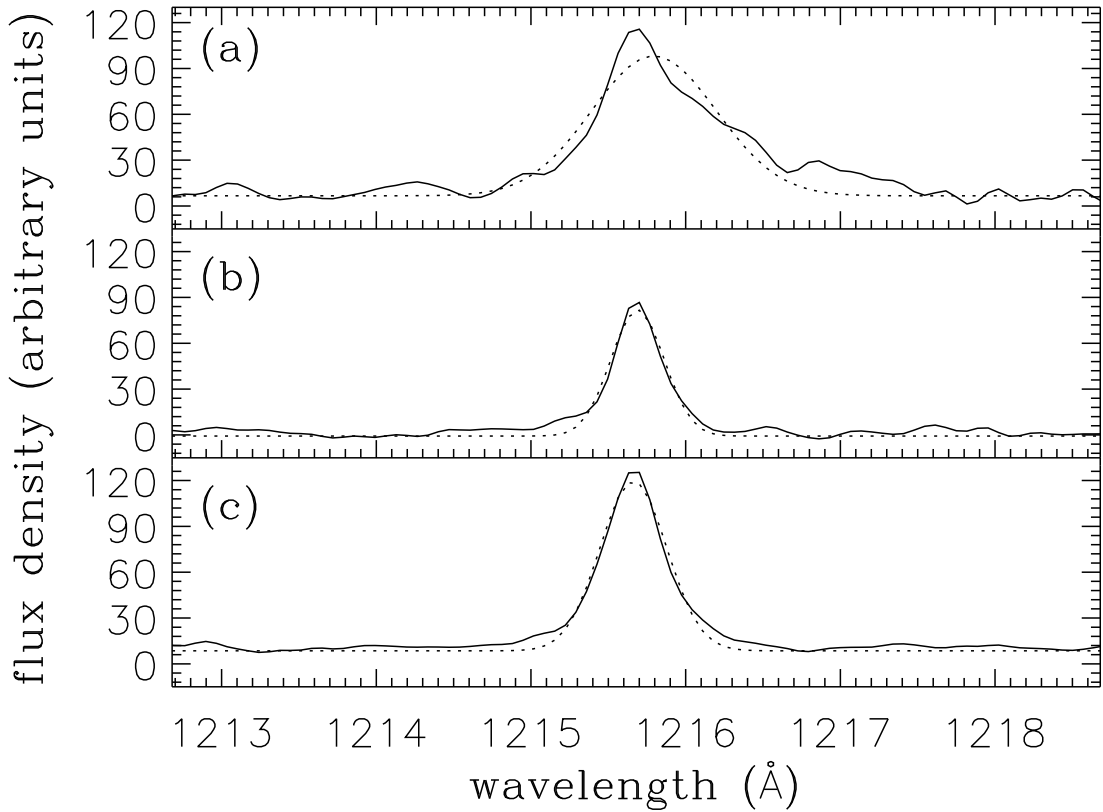}
\caption{\label{lineshapes.fig}
The Ly$\alpha$ candidate line shape for three averaged groups of spectra: (a) class 3 and 2 objects, (b) class 1 objects, (c) the 61 single-line DEEP2 objects that are known not to be LAEs.  Dotted lines show Gaussian fits.}
\end{figure}

Figure~\ref{lineshapes.fig} shows the averaged (composite) spectra, focusing on regions around the discovery line. Also shown are best-fitting Gaussian profiles.  The bottom panel (panel (c)) of the figure shows the composite spectrum of 61 single-line serendipitous DEEP2 objects that have been photometrically ruled out as LAEs (Kirby et al.\ 2007).  This low-$z$ composite spectrum can be used as a reference for comparing with the spectra of our LAE candidates.   The composite spectrum of our nine best LAE candidates --- i.e., class 3 and 2 objects --- is shown in Fig.~\ref{lineshapes.fig}(a).  It is not well fit by a Gaussian profile and instead exhibits the asymmetric line shape expected of \lya: the blue side has a steep drop while the red side has an extended tail.  This asymmetric shape is expected given that all the class 3 and 2 objects that make up this composite spectrum individually have asymmetric line profiles.  Overall, this composite spectrum further supports the \lya\ nature of the line in class 3 and 2 objects.  We present further analysis of this spectrum in \S~\ref{SpectralProperties.sec}. The composite spectrum of our class 1 objects, presented in Fig.~\ref{lineshapes.fig}(b), does not show asymmetry.  This lack of asymmetry argues against the class 1 objects being LAEs and while it is possible that some of the class 1 objects are LAEs, it is very likely that the class 1 sample is contaminated by low-$z$ interlopers. These results do not change with uniform rather than line weighting and we conclude that class 3 and 2 objects contain appreciable numbers of \lya\ emitters but that there may be significant foreground contamination among the class 1 objects.

\subsubsection{Presence or absence of other emission lines}\label{sec:extralines}

Our experience shows that the most likely single-line Ly$\alpha$ interlopers are H$\alpha$, H$\beta$, [\ion{O}{2}]3727, and [\ion{O}{3}]5007 emitters (see Martin \& Sawicki 2004; Martin et al.\ 2008). Objects with visible H$\alpha$ often show [\ion{N}{2}] or [\ion{S}{2}] lines.  Objects with visible H$\beta$ often show [\ion{O}{3}] and vice versa. If the discovery line is one of the likely interlopers then we might expect to see these other lines in the averaged spectra.  The averaged spectrum of the 61 non-LAE objects shows both [\ion{N}{2}] (6548\,\AA\ and 6583\,\AA) and the two [\ion{S}{2}] lines (6716\,\AA\ and 6731\,\AA) when we assume that the single line is H$\alpha$ (see Fig.~\ref{extralines.fig}).  With uniform weighting, [\ion{N}{2}] is not visible, and the significance of [\ion{S}{2}] decreases but that line is still detectable.  In contrast, no such lines are found in the composite spectra of our LAE candidates --- they are seen in neither the averaged spectrum of class 3 and 2 objects nor in that of the class 1 objects.   We also checked the composite spectra of our candidates for the presence of [\ion{O}{2}] 3727, [\ion{O}{3}] 4959, 5007, [\ion{S}{2}] 6716, 6730, H$\alpha$,  H$\beta$, and H$\gamma$ and none of these lines were found.  The lack of extraneous emission lines supports the claim that at least most of our candidates are \lya\ emitters.

\subsection{Redshift distribution}

We next investigate the contamination of our sample of high-$z$ LAE candidates by single emission line non-LAE galaxies (e.g., Kirby et al. 2007). We compare the redshift distribution of our LAE candidates against a sample of 61 single-emission line objects that are known {\it not\/} to be LAEs because they are detected in the $B$ and/or $R$ bands and clearly lack a spectral break across the emission line.  We do the comparison not in terms of a traditional redshift histogram, but rather in terms of the distributions of ``effective Ly$\alpha$ redshifts'' which are defined as \zeff = $(\lambda_{\rm em} / 1216~\rm{\AA}) - 1$. In other words, \zeff\ is the redshift that the object would have if it were a LAE.  The foreground interlopers in our LAE sample are presumably counterparts of the (known) low-$z$ single emission line objects, differing from them only in that they have a fainter continuum level such that they go undetected in the $B$ and $R$ bands.

\begin{figure}
\plotone{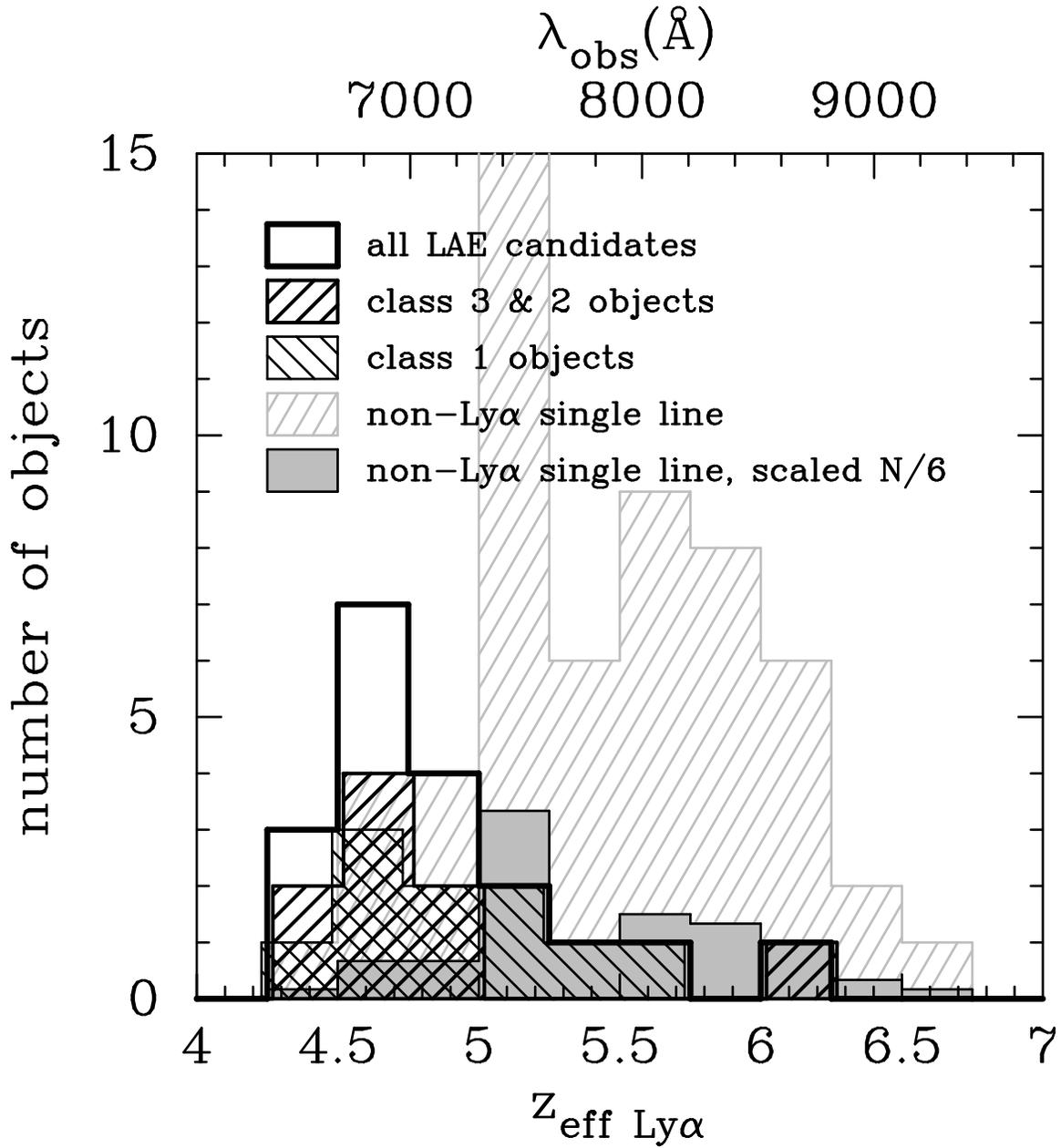}
\caption{\label{zhist.fig} 
Effective \lya\ redshift distributions --- i.e.,  redshift distributions assuming a rest-frame wavelength of 1216\,\AA\ for the single emission line in all cases.    For clarity some of the histograms have been slightly offset along the horizontal axis. The distributions of our LAEs are different from that of confirmed low-$z$ single-line non-LAE objects.}
\end{figure}

The bold histogram in Fig.~\ref{zhist.fig} shows the \zeff\ distribution for our sample of 19 class 3, 2, and 1 objects, while the hashed histograms are for class 1 (in one case) and for class 3 and 2 (in the other) objects only.  The light-colored histograms show the distribution of the known non-LAE objects, with the hashed histogram representing the actual distribution and the solid histogram showing that distribution scaled down by a factor of six.  

As expected, the sample of non-LAE objects is distributed more or less uniformly in \zeff\ across the available wavelength window.  Since the emission line in most of these non-LAE objects is expected to be H$\alpha$ (Kirby et al. 2007), it is not surprising perhaps that there are almost no objects with \zeff\ $< 4.397$, where $[\rm\lambda_{rest} (H\alpha) / 1216~\,\AA] - 1 = (6563~\,\AA / 1216~\,\AA) - 1 = 4.397$.  The rise in numbers from there to \zeff\ $\sim 5$ is likely a result of increasing volume probed, while the decrease beyond that is likely because of the higher density of night sky emission lines at redder wavelengths (Fig.~\ref{volume_vs_z.fig}) and the fall in instrumental efficiency at the longest wavelengths.  

In contrast to the non-\lya\ objects,  the distribution of our 19 LAE candidates is peaked at \zeff\ $\sim 4.7$ with an extended low-level tail out to the upper end of the available range.  The distribution of class 3 and 2 objects is strongly clustered around \zeff$\sim$4.7 and --- with the exception of object 3307.61 at $z$=6.0559 --- does not show a high-\zeff\ tail that would be consistent with the low-$z$ population.  The distribution of class 1 objects is also different from that of the non-\lya\ objects, although the class 1 objects do have a more prominent tail to higher redshifts.  It is thus quite plausible that some fraction of the class 1 objects are foreground interlopers. 

Scaling down the distribution of non-LAE objects by a factor of 6 causes it to roughly match the extended low-level, high-\zeff\ tail of the LAE candidate distribution beyond \zeff\ $\gtrsim 5$.  The factor of 6 is empirical and arbitrary because we have no {\it a priori} knowledge of the ratio of the number of non-LAE objects with detectable $B$- and/or $R$-band continuum to the number of non-LAE objects in which the continuum is too faint to be detectable in our imaging data.  If the above factor-of-six scaling is taken at face value, it would seem to indicate that most of the LAE candidates with \zeff\ $<5$ --- including a number of the class 1 objects at these \zeff\ --- are true LAEs, while the class 1 objects with \zeff\ $>5$ are mostly single emission line foreground galaxy contaminants.

We conclude that the \zeff\ distribution of our LAE sample is different from that of the confirmed low-$z$ single emission line objects, indicating that the LAE candidates are drawn from a different population than the low-$z$ objects.  Given the additional evidence in the form of the shapes of the emission lines, we conclude that our class 3 and 2 LAE candidates are indeed high-$z$ galaxies, while the class 1 candidates may contain a significant fraction of low-$z$ interlopers, although it is impossible at this point to tell which individual class 1 objects are LAEs and which ones are interlopers. 

\subsection{Spectroscopic follow-up}

\begin{figure}
\plotone{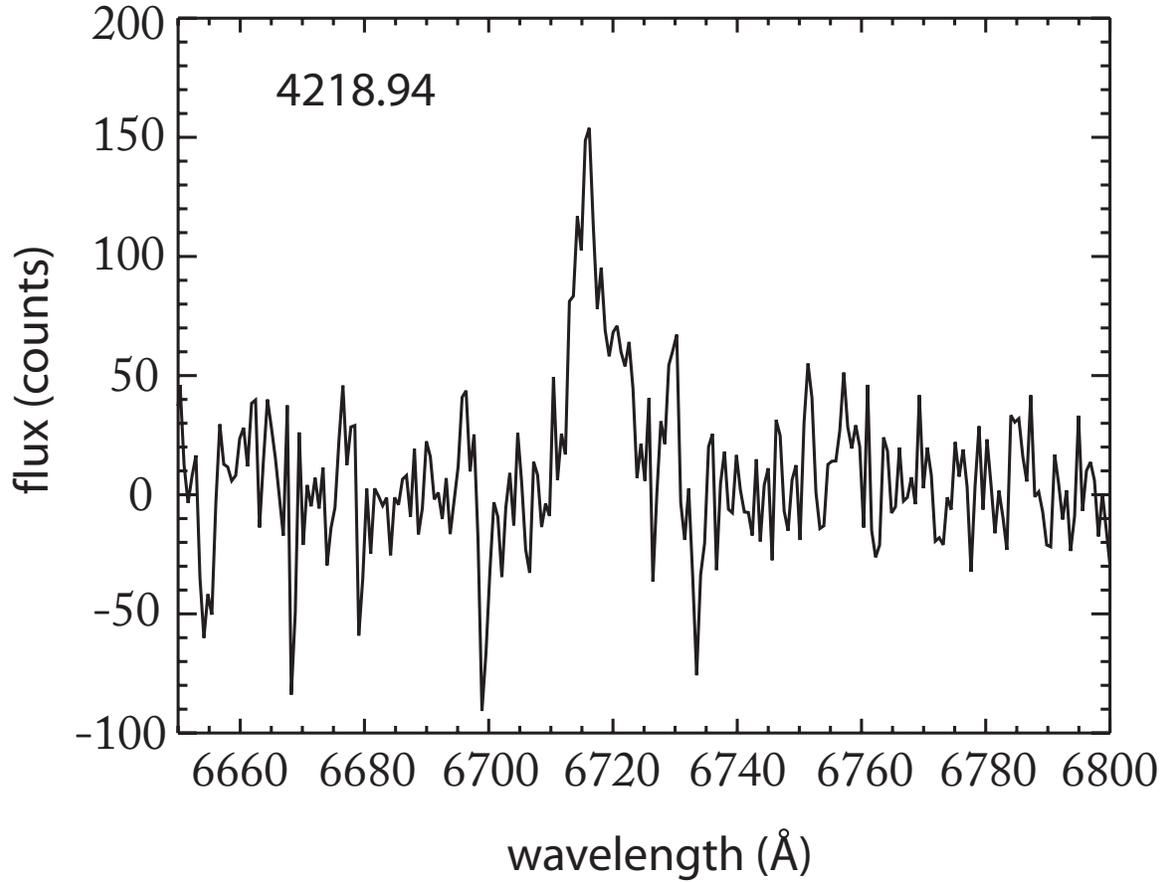}
\caption{\label{4218.094.followup.fig} 
Follow-up DEIMOS spectrum of object 4218.94 taken with the 600 line grating at $R\sim$ 2000 (FWHM of 3.5\AA) and with an exposure time of 3 hours.  The spectral region around the candidate line is shown.  The asymmetric shape expected of \lya\ emission can be clearly seen.}
\end{figure}

Finally, we have also re-observed one of our candidates, namely the class 3 object 4218.94, using DEIMOS at lower dispersion and with an exposure time of 3 hours. Figure~\ref{4218.094.followup.fig} shows the region around the \lya\ in these followup data.  This spectrum clearly shows the asymmetric line profile shape expected of \lya, giving us further confidence that this object is a $z$=4.53 LAE.  

\vspace{3mm}

In summary, we have high confidence that the bulk of our class 3 and class 2 objects are genuine, high-$z$ LAEs because  both individually and as an ensemble they pass the full range of tests that we have subjected them to.  At the same time there is a significant probability that the class 1 candidates are contaminated by foreground interlopers:  their individual and composite line shapes are not asymmetric but resemble those of known low-$z$ galaxies, but, on the other hand, their redshift distribution is different from that of single-line low-$z$ objects, arguing that a significant number --- perhaps around half --- of class 1 objects could be LAEs;  the absence from our composite spectra of emission lines expected of low-$z$ interlopers (\nii, \sii, etc) does not confirm our class 1 objects as LAEs, but does instead keep them as a viable candidate population.   

We conclude that we have high confidence that the bulk of our class 3 and 2 objects are genuine high-$z$ LAEs, while there is significant probability that the class 1 objects are contaminated by foreground interlopers.  In the following sections we use these samples of objects --- relying mainly on the much higher quality class 3 and 2 objects --- to constrain the number density of high-$z$ LAEs and to study the physical properties in these objects from the \lya\ emission line profile.

\section{NUMBER DENSITY OF LYMAN ALPHA EMITTERS AT z$\sim$4.7} \label{numberdensity.sec}

\subsection{The simple approach}

\begin{figure}
\plotone{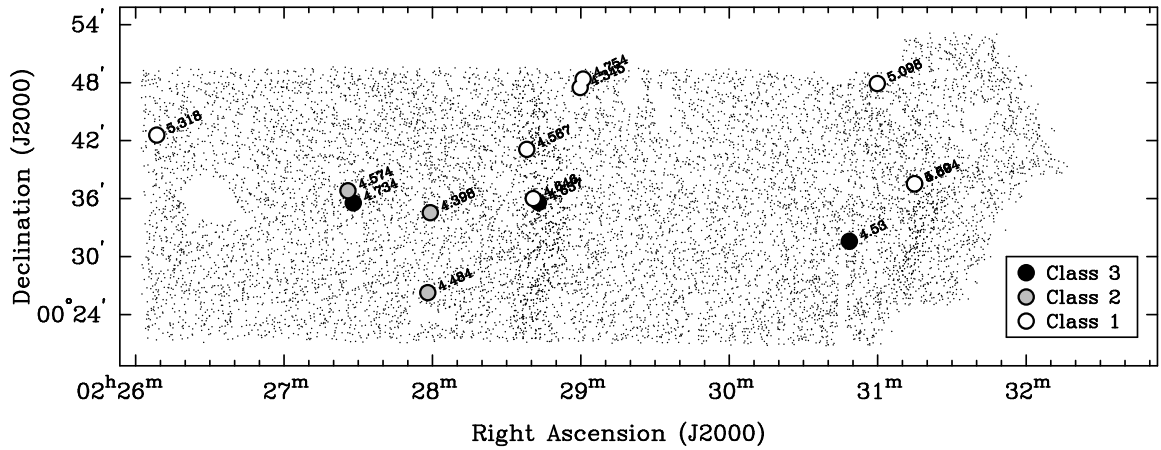}
\caption{\label{positions.fig} 
Positions and \lya\ redshifts of our LAE candidates in the 4th DEEP2 field.  Small points show the positions of the 11,250 DEEP2 spectroscopic targets in this field and illustrate the topology of the survey.  It is clear that our LAE candidates are clustered both is space and in redshift.}\end{figure}

A large volume free of OH sky-lines is accessible to us in the redshift interval $z$=4.4--4.9 (Fig.~\ref{volume_vs_z.fig}) and, in fact, all but one of our nine highest-quality (class 3 and 2) objects are found in this redshift range (Fig.~\ref{flux-vs-wav.fig}).  High-$z$ LAEs are a strongly clustered population (e.g., Ouchi et al.\ 2003), and we see signs of clustering in our own data: Fig.~\ref{positions.fig} illustrates that our \lya\ candidates are highly clustered on the sky with large parts of the field completely void of them.  The concentration of redshifts around $z$$\sim$4.7 could thus be a result of large-scale structure.   However, it is also the case that our survey is by far most sensitive to finding LAEs around \zs4.7 because of the combination of the large differential volume, dearth of strong skylines at the corresponding wavelengths (Fig.~\ref{volume_vs_z.fig}), and fainter limiting luminosity than is reached at higher redshifts.  While clustering of LAEs is certainly important in estimating their luminosity function, we defer the analysis of cosmic variance effects until our sample includes fully all four of the DEEP2 fields. For now, we take the observed numbers of objects at face value and proceed with the LF analysis, noting only that our results are based largely on a single field (Field 4 of DEEP2), although one that spans a large volume.

The left axis of Fig.~\ref{LF.fig} shows the cumulative number counts of our objects.  The open circles show all the class 3 objects within \zlya=4.4--4.9, while the filled circles also include the class 2 and 1 objects. The latter are therefore likely to be contaminated by foreground interlopers that possibly lurk among the class 1 objects as well as the potentially lensed and/or ISM-attenuated class 2 objects.  The filled circles thus show the more conservative estimate and henceforth we focus on this sample.  Note that the DEEP2 points shown in Fig.~\ref{LF.fig} are lower limits on the true number density of LAEs for the following two reasons.  First, our by-eye search no doubt missed some objects, particularly at the faint end of the population.  Second, our objects' luminosities are strictly speaking lower limits because lacking detailed positional information we have no way of correcting for flux slit losses on an object-by-object basis;  moreover, some of our objects were observed in non-photometric conditions further making their fluxes lower limits.  These two effects cause our number counts to be lower limits (the first limit is in the number direction, the second is in the luminosity direction), with the area above and to the right of our points in Fig.~\ref{LF.fig} being permitted by the data.

To translate from the cumulative number distribution for our survey (left-hand axis of Fig.~\ref{LF.fig}) to a cumulative number {\it density} distribution (right-hand axis) we need to divide by the survey volume. Here we use the volume that corresponds to integrating over the comoving distance from $z$$=$4.4 to $z$$=$4.9 multiplied by the area subtended by the searched slits as a function of redshift (i.e., the integral, from $z$$=$4.4 to $z$$=$4.9, under the lower curve in Fig.~\ref{volume_vs_z.fig}). This operation gives a volume of $2.89\times10^4$~Mpc$^3$.  Our choice of $\Delta z$ is motivated by the fact that almost all of our class 3 and 2 LAEs lie in the $z$$=$4.4--4.9 interval, and by the fact that the bulk of OH skyline-free survey volume lies in that redshift range (Fig.~\ref{volume_vs_z.fig}).  Adopting a larger redshift range would decrease the number density.  In the extreme that allows the full (but skyline-reduced) survey volume from the hard limits of $z$=4.2 to $z$=6.6, the volume would increase by a factor of 2.4 from the $2.89\times10^4$~Mpc$^3$ that we adopted to $6.92\times10^4$~Mpc$^3$; this effect would result in formally lowering the DEEP2 number density shown in Fig.~\ref{LF.fig} by a corresponding factor of 2.4. In practice, however, the high-$z$ end of this longer redshift range would sample only the most luminous and rare objects, and thus the cumulative number densities are unlikely to be lower by as much as that factor of 2.4 from those calculated using the more restricted $z$$=$4.4--4.9 interval.  

Our assumption about the survey volume here is inaccurate for the following reason.  Intrinsically luminous objects located {\it outside} our slits but sufficiently close to them could potentially contribute to the observed counts by virtue of flux spilling over into the slit; this effect could, in principle, lower our number densities. On the other hand, intrinsically faint objects will drop out of our sample if they are located sufficiently close to the edge of the slit to lose significant amounts of their flux due to slit losses; this effect acts in the opposite direction, namely to increase the true number density.  These effects are similar to one that affects LF estimates in narrowband imaging surveys.  There, most narrowband filter transmission curves are not perfect top-hats but are better approximated by Gaussians, and, consequently, in the absence of a precise spectroscopic redshift, an object's true flux cannot be determined because its position with respect to the filter profile is unknown. This uncertainty propagates into a bias in luminosity function estimates (further compounded by any redshift-space clustering that preferentially favors one part of the filter transmission curve over the rest), just as our uncertainty of the object's precise position in the slit propagates into a similar uncertainty in our simple determination of the LF.   The effect can be corrected on a statistical basis in both imaging (Shimasaku et al.\ 2006) and spectroscopic surveys (Martin \& Sawicki 2004; Martin et al.\ 2008), but many narrowband imaging surveys do not make this statistical correction, assuming instead a top-hat filter transmission curve. In the present section we take the similar approach of assuming a top-hat slit transmission function, and we defer the more accurate approach until \S~\ref{slitlosscorrection.sec}. 

We compare our straightforward number counts to those from the large narrow-band-selected LAE samples of Dawson et al.\ (2008) and of the Subaru Deep Survey (SDS; Ouchi et al.\ 2003; Shimasaku et al.\ 2006; Kashikawa et al.\ 2006), which contain some of the largest samples of  LAEs published to date.  We also show the recent results from the dedicated Magellan $z$=5.7 multislit spectroscopic search by Martin et al.\ (2008).  For the Ouchi et al.\ (2003) results, we show two curves: the upper curve shows the number counts of narrowband-excess objects selected using a simple EW cut similar to that of Hu et al.\ (1998); the lower curve is for their sample after it has been reduced by a color-color cut correction that eliminates objects without a significant Lyman break, as well as a statistical correction for foreground interlopers based on Monte Carlo simulations (see Ouchi et al.\ 2003).  

Our adopted redshift interval $z$$=$4.4--4.9 is most directly comparable with the $z$$\sim$4.5 results of Dawson et al.\ (2008) and the $z$$\sim$4.9 results of Ouchi et al.\ (2003).  The number counts of objects of all three of our confidence classes are in good agreement with the results of these two narrowband imaging surveys.  If we consider our class 3 objects only then the DEEP2 number densities are a factor of a few below those of the narrowband imaging surves.  This apparent discrepancy is entirely expected, however, because our by-eye search is less than 100\% complete and the flux estimates of our LAEs suffer from (unknown and unknowable) slit losses that may make them appear fainter than they really are.  Overall, we conclude that our serendipitous spectroscopic technique is effective at finding samples of high-$z$ LAEs.

\subsection{Accounting for slit flux losses}\label{slitlosscorrection.sec}

In this section we attempt to estimate the LAE luminosity function by properly correcting for flux slit losses in a statistical way.  We use the statistical LF estimation technique developed for this purpose by Martin \& Sawicki (2004).  The technique is described in detail in Martin \& Sawicki (2004) and here we review it only briefly.

The approach is to make a set of parametric models of the underlying LAE population and to simulate the observational and instrumental processing of these model populations to arrive at a predicted number of objects for each set of parameters.  A comparison of these predictions with the observed number of objects (or, in our case, lower limit on the number counts, given the intrinsic incompleteness of our by-eye selection), then constrains the parameters of the underlying luminosity function. 

We assume an LF described by the Schechter (1976) function, 
\begin{equation}\label{schechter.eq} 
\phi(L/L^*)d(L/L^*) = \phi^* (L/L^*)^{\alpha} e^{-L/L^*} d(L/L^*),
\end{equation}
with freely-varying number density (\phistar) and characteristic luminosity (\Lstar) but with a fixed faint-end slope (we adopt two values: $\alpha$=$-$1.2 and $\alpha$=$-$1.6). In principle the faint-end slope could also be a free parameter, but here we have fixed it given that our DEEP2 survey does not probe sufficiently deep to constrain the faint end. 

For a given set of Schechter function parameters, we expect our survey to find the following number of objects:
\begin{equation}\label{expectedN.eq}
N(\phi^*, L^*, \alpha) = \int^{\infty}_{L_{min}} \phi(L/L^*) V(L/L^*)
\xi(L/L^*) d(L/L^*).
\end{equation}
We set the limiting luminosity $L_{min}$ based on the limiting flux, which we set to $f_{min}$=3$\times$$10^{-18}$ as this contains most --- but not all --- of the objects in our catalog. Given the subjective, by-eye nature of our search, we do not know the survey completeness, $\xi$; here we set it equal to unity thereby guaranteeing that the expected number of objects, $N$,  is a hard upper limit.  The survey volume $V$ is luminosity-dependent since it decreases for less luminous objects as intrinsically faint objects need to be closer to the middle of the slit to ensure that they have sufficient flux after slit losses to make it above the detection threshold.  As in Martin \& Sawicki (2004), we compute the luminosity-dependent survey volume from the cosmological volume increment per unit redshift, the accessible redshift interval (which we set as $z$$=$4.4--4.9), the effective slit area of the present survey (21.30~\sqam), and the slit loss function defined for our 1\arcsec\ slits and 0.8\arcsec\ Gaussian seeing.

\begin{figure}
\includegraphics[width=10cm]{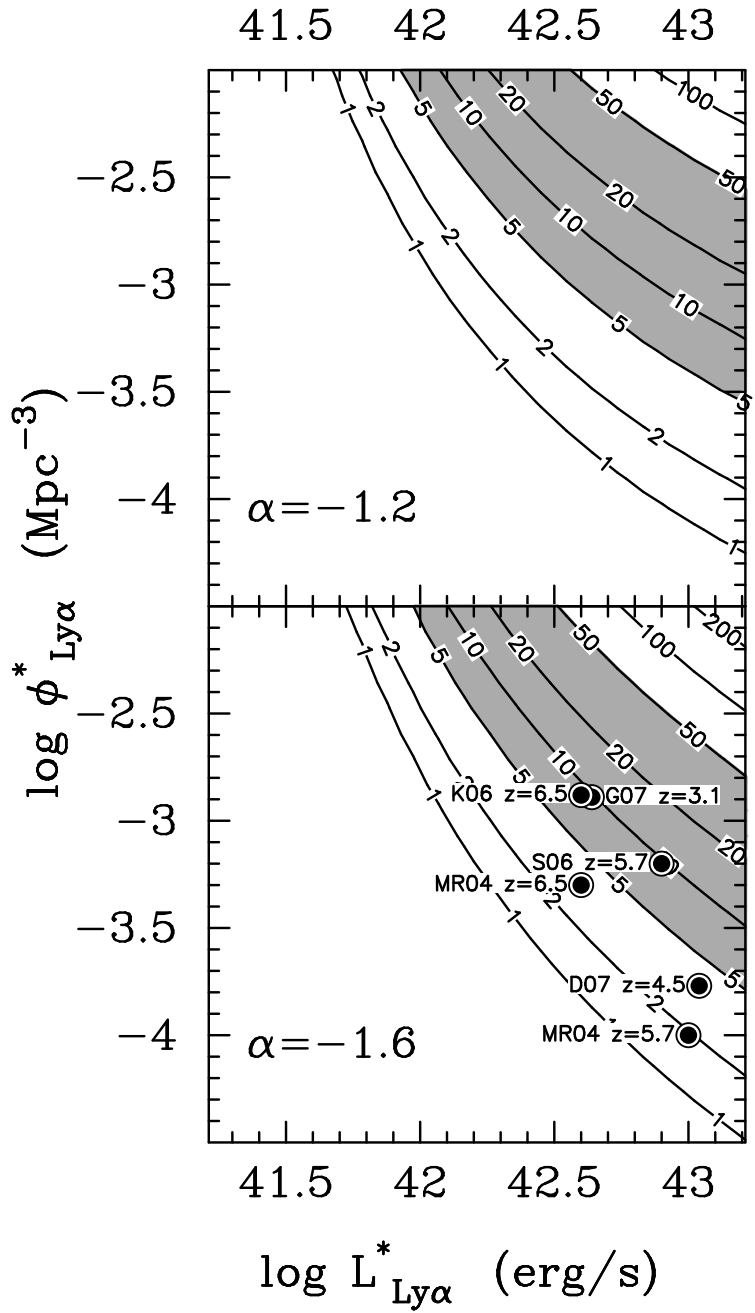}
\caption{\label{Nexpected.fig} 
Constraints on the luminosity function parameters. The expected number of objects brighter than 3$\times$10$^{-18}$ erg~s$^{-1}$ as a function of the Schechter parameters of the LF. The two panels assume different fixed faint-end $\alpha$, but, as can be seen from the small differences between the two panels, the choice of $\alpha$ is not dramatically important.   The shaded region shows the values of the Schechter parameters that are permitted by the number of objects we detected (the upper limit of this region assumes that we are missing no more than 9 out of 10 objects).  In the bottom panel we show the Schechter fit parameters reported by a number of recent \lya\ surveys: Gronwall et al.\ (2007, G07), Malhotra \& Rhoads (2004, MR04), Shimasaku et al.\ (2006, S06), Kashikawa et al.\ (2006, K06), and Dawson et al.\ (2007, D07).}
\end{figure}

The contours in Fig.~\ref{Nexpected.fig} show the expected number of $z$$=$4.4--4.9 LAEs brighter than 3$\times$10$^{-18}$\cgsflux\ for different combinations of Schechter \phistar\ and \Lstar\ (with $\alpha$ fixed at $-$1.2 and $-$1.6).  There are five class 3 LAEs above that flux limit in our survey, which --- given that our survey is incomplete --- gives a hard lower limit on the true number of LAEs.  Thus, the shaded region of parameter space to the upper right of the ``5" curve is permitted by our observations. Although our current catalog is certainly incomplete, it seems unlikely that we miss more than nine out of ten true LAEs, and so we set a somewhat arbitrary but illustrative upper limit of $<$50 LAEs in our plot.  The shaded regions in Fig.~\ref{Nexpected.fig} shows the regions of parameter space permitted by these two bounds.  Note that there difference between the results for two faint-end $\alpha$ are very minor --- a consequence of he fact that at the depths reached by our study we are just starting to probe into the faint end of the LAE LF. 

The Schechter fit results reported by a number of recent \lya\ surveys are plotted in the bottom panel of Fig~\ref{Nexpected.fig}. The results shown are those of Gronwall et al.\ (2007) at \ze3.1, Malhotra \& Rhoads (2004) at \ze5.7 and 6.5, Shimasaku et al.\ (2006) at \ze5.7, Kashikawa et al.\ (2006) at \ze6.5, and Dawson et al.\ (2007) at \ze4.5.  Although different authors assume (or measure) different values of $\alpha$ in their fitting, most of these $\alpha$ are close to $\alpha$= $-1.6$ and, at any rate, the differences in our derived LF parameters are not strongly dependent on $\alpha$.   We are thus justified in plotting the results of these diverse studies together in one common, $\alpha$=$-1.6$, panel without explicitly adjusting for differences in their $\alpha$ values.  

When inspecting Fig.~\ref{Nexpected.fig} one should note that, in contrast to many narrowband imaging surveys, our spectroscopic survey has no equivalent width detection threshold and therefore in principle, and barring issues to do with our survey completeness, is capable of detecting \emph{all} LAEs down to a limiting line flux. In contrast, some of the narrowband surveys do impose an equivalent width cut, and as a result would miss some of the objects that our data are sensitive to. 

It is reassuring to note that our DEEP2 results are in reasonable agreement with Schechter function fits reported by other LAE surveys at similar and higher redshifts.  Overall, then, the agreement between our results and those of narrowband imaging surveys that is illustrated in Fig~\ref{Nexpected.fig}  bodes well for the future of multi-slit spectroscopic searches as a technique for studying the LAE population.

\section{THE SPECTRAL SHAPE  OF THE \lya\ LINE}\label{SpectralProperties.sec}

The spectral profile of the \lya\ line carries information about the conditions in and around the galaxy that emits it.  Here we fit the stacked, composite spectrum of our nine class 3 and 2 objects (\S~\ref{sec:lineshapes}) with simple models that attempt to model these conditions.  Our composite spectrum consists of only nine individual spectra of relatively poor S/N, so we do not attempt to fit sophisticated models such as those of Hansen \& Oh (2006) or Verhamme, Schaerer, \& Maselli (2006), but rather use simple Gaussian toy models instead. 

One strong limitation of fitting stacked \lya\ spectra which we wish to point out at the outset is that spectral averaging relies on shifting the individual spectra to a common rest frame. This rest frame can be determined only imperfectly for each individual object because of the stochastic nature of the IGM absorption of \lya\ photons that truncates the blue wing of the line.  Consequently, the stacked spectrum suffers some unknown degree of "smearing" which cannot be accounted for in a simple analysis such as that presented here. Although this effect limits the usefulness of the analysis, we press ahead in hope of gaining some, even if imperfect, insights.

The first of our toy models is a single emission line that is meant to replicate emitting gas within a high-$z$ galaxy that's broadened by large-scale motions within the object.  The second model is a two-component model that adds a second \lya\ line redshifted with respect to the main line.  The physical motivation for the two-component model is that of backscattering of \lya\ photons from an outflowing galactic wind or expanding gas shell that's powered by the galaxy's star-forming activity (e.g., Dawson et al.\ 2002; Mas-Hesse et al.\ 2003; Ahn 2004; Westra et al.\ 2005; Hansen \& Oh, 2006; Kashikawa et al.\ 2006).  

Both models consist of a Gaussian-profile ``main" \lya\ line with a width, height, and central wavelength that are free parameters of the fit.  In the two-component model, the second, back-scattering line is likewise modeled using a Gaussian profile with a width, height, and central wavelength that are free parameters. The velocity offset between the main and backscattering lines is also a free parameter of the two-component model.  In both models we mimic the absorbing effect of the intervening IGM by setting to zero all flux blueward of the centre of the main \lya\ line.  We then convolve the resulting spectra with a Gaussian smoothing kernel to simulate the effects of the instrumental point spread function (PSF).  Since we don't know the sizes and in-slit placement of the objects that make our composite spectrum, we cannot determine the instrumental PSF from the data by using, e.g., night sky-lines profiles.  Instead, we also make the instrumental PSF FWHM a free parameter of the fit. Once we have generated a large set of model spectra that reflect a large grid of parameter combinations, we then find the best-fitting models by means of $\chi$$^2$ minimization.

\begin{figure}
\plotone{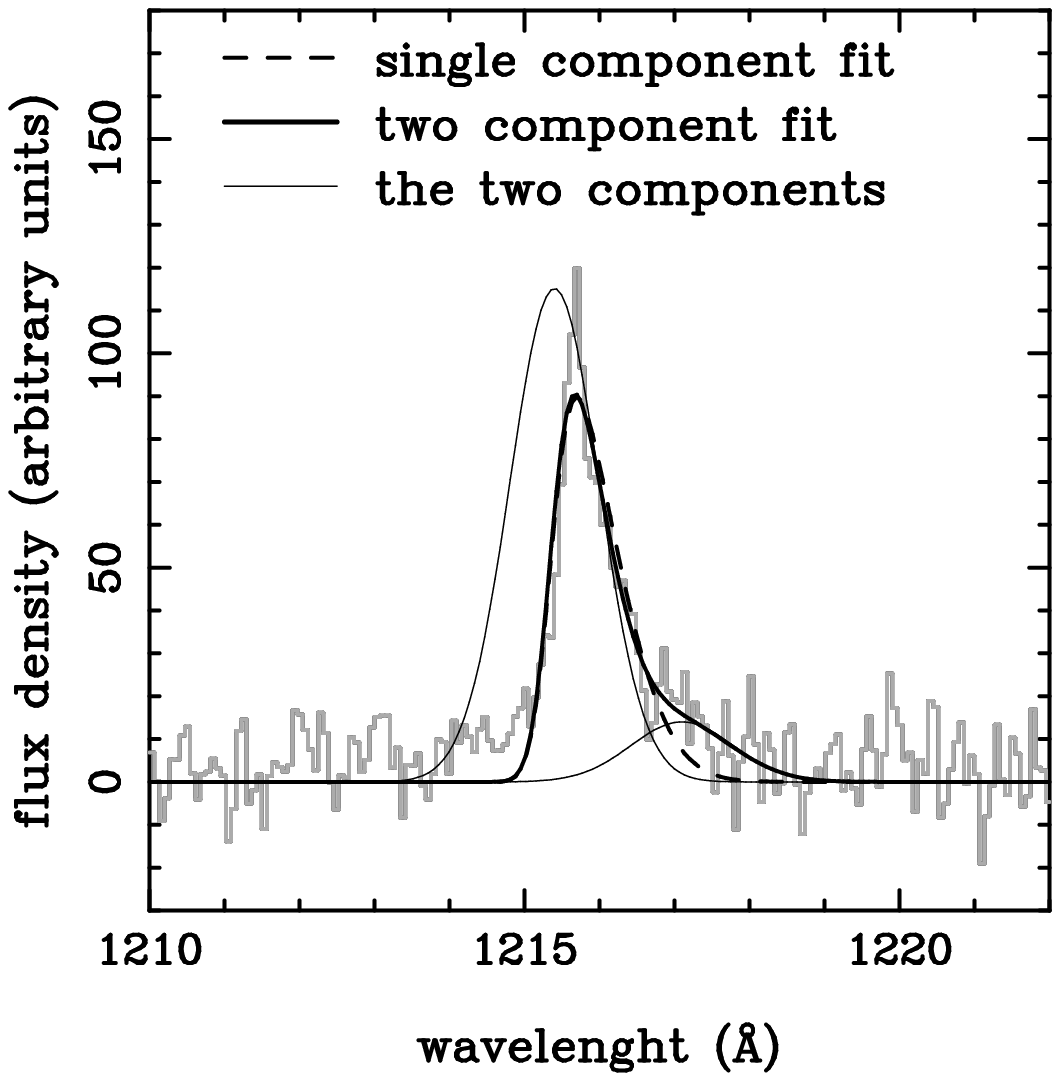}
\caption{\label{line_fit.fig} 
Model fits of the composite spectrum of our class 3 and 2 objects.  The thin solid lines show the two components of the two-component model as they would appear in the absence of intergalactic absorption. }
\end{figure}

Figure~\ref{line_fit.fig} shows a comparison of the best-fitting models with the data. The observed spectrum is the same composite as that shown in Fig.~\ref{lineshapes.fig}(b) but unlike in Fig.~\ref{lineshapes.fig}, here the data have not been smoothed.  The dashed line shows the best-fitting single-component fit. The underlying (i.e., not truncated) \lya\ line profile here has a FWHM of 1.7\,\AA, which corresponds to a velocity dispersion of $\sigma_v$=178~\kmpersec.  This model, which simulates the \lya\ line as a Gaussian truncated in the blue by intervening intergalactic gas, accounts well for the observed asymmetric \lya\ line shapes (e.g., Dawson et al.\ 2002; Hu et al.\ 2004). The width of the line is caused by large-scale motions of gas inside the galaxy or possibly by extragalactic outflows, while its truncated blue side is due to absorption by intervening neutral gas along the line of sight. The velocity dispersion of 178~\kmpersec\ in our best-fitting single-component model is comparable to the 200~\kmpersec\ found in the composite \lya\ spectra at \zs 5.7 by Hu et al.\ (2004). 

A failing of the single-component model is that it underpredicts the flux at $\sim$1217\,\AA.  This excess flux could be the signature of back-scattering of \lya\ photons off of a galactic outflow (e.g., Dawson et al.\ 2002; Mas-Hesse et al.\ 2003; Ahn 2004; Westra et al.\ 2005; Hansen \& Oh, 2006; Kashikawa et al.\ 2006). Our two-component model aims to simulate this galactic outflow scenario. The best-fitting two-component solution is shown in Fig.~\ref{line_fit.fig} with the thick solid line. It gives a better match to the data at $\sim$1217\,\AA\ than does the single-component model, while providing an equally good fit in other regions of the spectrum.  In the best-fitting two component model, the parameters of the main \lya\ line are little different from those of the single-component model, including a FWHM of 1.40\,\AA, which corresponds to $\sigma_v$=147~\kmpersec.  The velocity offset between the main \lya\ line and the outflow line corresponds to 420~\kmpersec\ --- an offset that is similar to those found in other \lya\ emitters (e.g., Dawson et al.\ 2002; Westra et al.\ 2005; Kashikawa et al.\  2006). At the same time, the width of the outflowing line ($\sigma_v$=167~\kmpersec) is lower than the 260~\kmpersec\ of Kashikawa et al.\ (2006) at \zs6.5, and in the single lower-$z$ objects analysed by Dawson et al.\ (2002; 320~\kmpersec) and Westra et al.\ (2005; $\sim$405~\kmpersec).  In low-$z$ galaxies, outflow speeds correlate with galactic escape velocity (Martin 2005), so these velocity differences may suggest that the \lya\ emitters in our sample are lower-mass objects than those studied by the aforementioned authors.  However, while it is tempting to equate the observed velocity offsets with outflow speeds, the relation between these two quantities is probably less straightforward (see, e.g., the models of Hansen \& Oh, 2006);  unfortunately, at the moment our data are not good enough to allow more detailed modeling of the properties of the outflows. 

Fig~\ref{line_fit.fig} also shows a possible flux excess to the blue of the main \lya\ line. We note that if this excess is real then it could be due to some \lya\ photons escaping absorption by the intervening material akin to the escape of ionizing continuum radiation (e.g., Shapley et al.\ 2006; Iwata et al.\ 2008).  If real, the escape fraction of \lya\ photons in the blue wing of the line would be $\sim$15\%.  It will be interesting to investigate this issue with the full DEEP2 LAE dataset.

Overall, we conclude that the two-component fit that attempts to model a backsattering of \lya\ photons from an outflowing galactic wind gives a good match to the observed composite spectrum.  We regard this result as consistent with the idea of large-scale outflows of material in high-$z$ galaxies.

\section{DISCUSSION AND SUMMARY}\label{discussion.sec}

Deep LAE surveys are an important tool for the study of formation and evolution of galaxies. Despite their many very important successes, narrowband imaging surveys for LAEs have their limitations:  the numbers of spectroscopically-confirmed objects are still modest and the lack of detailed spectroscopy limits the ability to study in detail such properties as galactic outflows. Moreover,  the lack of spectroscopic redshifts for all the LAEs introduces biases in luminosity function and correlation function studies.  Thus, while we must continue to press with narrowband imaging surveys, we must also develop other, complementary  techniques.  Martin et al.\ (2008) compare in detail the merits of spectroscopic and narrowband imaging imaging LAE surveys and discuss the future prospects for these techniques in the coming era of extremely large telescopes; here, we focus more narrowly on the near-term prospects of exploiting the remaining $\sim$80\% of the DEEP2 spectroscopic database. 

In this paper we have demonstrated that a large blank-sky spectroscopic survey can find significant numbers of LAEs.  Blank-sky spectroscopic searches have been attempted in the past (e.g., Crampton \& Lilly 1999; Martin \& Sawicki 2004; Tran et al.\ 2004; Martin et al.\ 2008), although the number of LAEs fund by them to date is very small.   Here we have piggy-backed on a large intermediate-$z$ survey that is DEEP2 and in just $\sim$20\% of these data we have found 9 objects that are bona fide LAEs (and have a further 10 of which some may also be such).  

So far, ours is a modest sample, but we have been able to use it to show that spectroscopic LAE searches can do well at constraining properties of the population.  Specifically, we found that the number density of objects in our survey is consistent with those found by narrowband imaging searches (\S~\ref{numberdensity.sec}), and --- using only our stacked discovery spectra with no additional follow-up data, we confirmed the interpretation that LAEs are associated with large-scale outflows of material (\S~\ref{SpectralProperties.sec}). Much more could be done with a bigger sample and with deep follow-up spectroscopy. 

\begin{figure}
\plotone{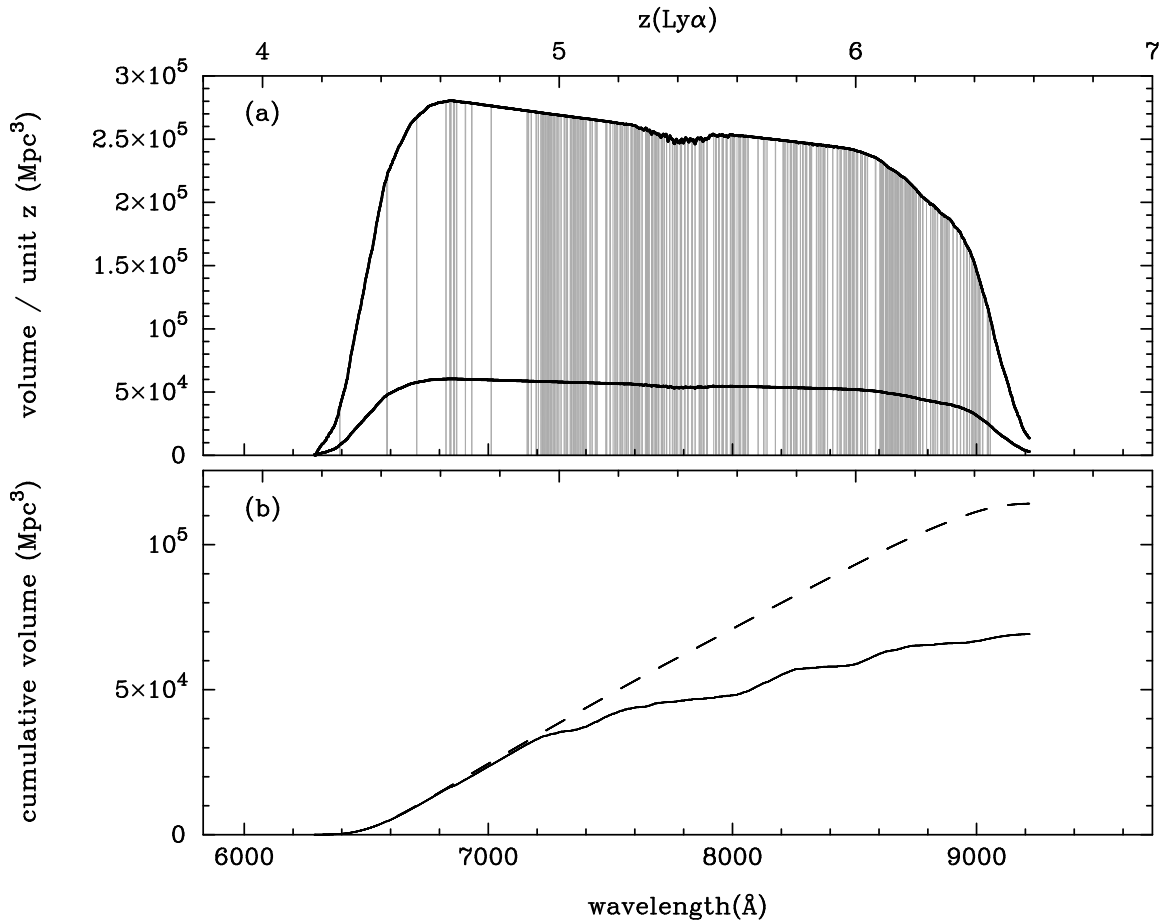}
\caption{\label{volume_vs_z.fig} 
The wavelength coverage of DEEP2, and the survey's volume as a LAE survey.  The {\it top panel} shows the volume surveyed in the present work which is based on 83 masks  (lower thick line) and that  covered by the full DEEP2 survey of 385 masks (upper thick line).  The volume shown is that subtended by the full DEEP2 slits --- i.e., no correction for obscuration by low-$z$ galaxies or slit-end effects have been applied. Note that the volume is shown per unit redshift ($\Delta z$=1).  The vertical gray lines mark regions of high sky noise due to atmospheric emission lines. The {\it lower panel} shows the cumulative volume of the present survey, with the dashed curve showing the volume without correcting for wavelength regions made unusable by night skylines and the solid curve showing the volume after applying such a correction. While there are many skylines in the red, DEEP2's spectral resolution provides enough clear inter-skyline regions to give significant volume for LAE searches at higher redshifts.}
\end{figure}

\begin{figure}
\plotone{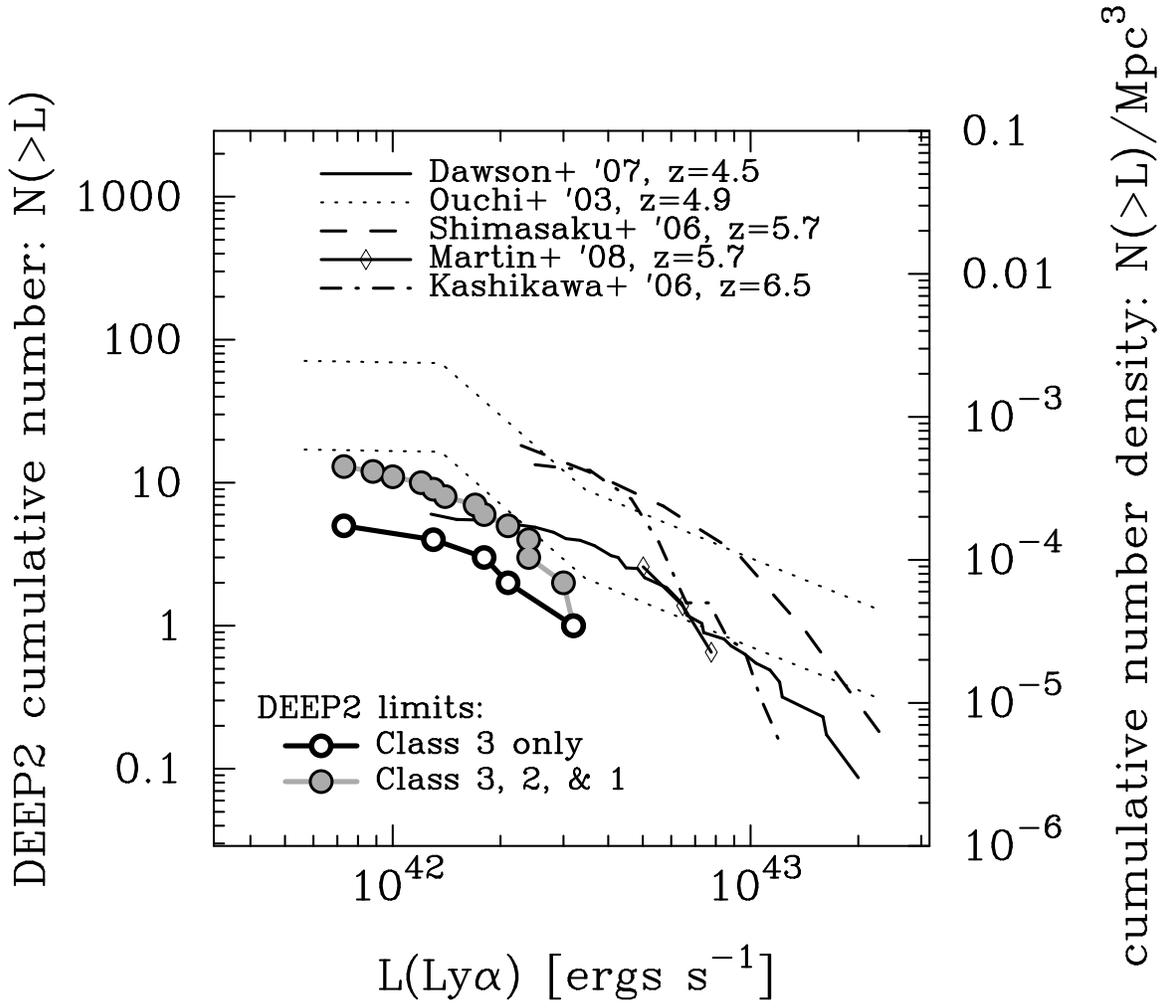}
\caption{\label{LF.fig} Cumulative number distributions. The left  axis shows the lower limit on the cumulative number of $z$=4.2--4.9 DEEP2 LAEs.  The points shown are a lower limit in both number and flux, as described in the text.  The right axis shows the cumulative number density distribution for our DEEP2 LAEs as well as for some representative recent surveys.
}\end{figure}

A simple extension of the by-eye search described in this paper to the remaining $\sim$80\% of the DEEP2 data can be expected to yield five times more LAEs in total, i.e. 42 class 3 and 2 objects (or 88 if class 1 objects are also included).  However, further improvements in efficiency over that achieved here are possible.  Specifically, in the present by-eye search we are sure to have missed many objects since the human eye is not an optimal tool when confronted with vast amounts of data.  An automated search of the available dataset should be much more efficient.  Our first experiments with implementing such an automated search are very encouraging and show that we can recover the objects found by eye as well as find additional, missed line emitters.  We speculate that with an automated search we should be able to double the number of LAEs we find per unit area --- as is also suggested by a comparison of our number counts with those of narrowband imaging surveys (Fig.~\ref{LF.fig}).  An automated search of DEEP2 would thus give us a sample of $\sim$90 class 3 and 2 objects, all with spectroscopic redshifts and with high-resolution spectra suitable for the study of the details of gas kinematics.  Finally, the spectral coverage of DEEP2 gives us the potential to probe LAEs all the way up to $z$$\sim$6.6 (Fig.~\ref{volume_vs_z.fig}), thus allowing us to study the evolution of the population from near the epoch of reionization to $z$$\sim$4.5 using a single, uniform dataset. Another clear advantage of an automated search is that it would allow us to accurately calibrate the detection efficiency through simulations and thus correct for it in our LF analysis.  Moreover, analyzing the full four-field DEEP2 dataset would also allow us to constrain the importance of field-to-field variance.  Clustering effects  appear important even in large fields such as those from Subaru/Suprimecam and constraining them using the four DEEP2 fields should be a worthwhile endeavor. 

In summary, we have shown that the systematic examination of a large spectroscopic database can yield significant numbers of faint high-$z$ line emitters which are immediately suitable for further studies such as the determination of their number density (\S~\ref{numberdensity.sec}) or kinematics (\S~\ref{SpectralProperties.sec}). Automating such a search and applying it to an even larger spectroscopic database should be possible and should yield a large sample of high-quality high-$z$ LAEs suitable for a variety of studies.

\acknowledgements
We wish to recognize and acknowledge the very significant cultural role and reverence that the summit of Mauna Kea has always had within the indigenous Hawaiian community; we are most fortunate to have the opportunity to conduct observations from this mountain. We thank Alison Coil and Sandy Faber for useful discussions, and all the members of the DEEP2 team for their contributions to making possible the data used here.  We thank Jerzy Sawicki for a careful reading of the manuscript and many useful comments. Parts of the analysis presented here made use of the Perl Data Language (PDL) that has been developed by K.\ Glazebrook, J.\ Brinchmann, J.\ Cerney, C.\ DeForest, D.\ Hunt, T.\ Jenness, T.\ Luka, R.\ Schwebel, and C. Soeller,  which can be obtained from http://pdl.perl.org. PDL provides a high-level numerical functionality for the perl scripting language (Glazebrook \& Economou, 1997).  This work was supported in part by a Discovery Grant from the Natural Sciences and Engineering Research Council of Canada, NSF grants AST-0071198 and AST-0507483 funding from the NASA/STScI and the Canadian Space Agency, and by the David and Lucile Packard Foundation.

\clearpage

\begin{deluxetable}{lcccccccc}
\tablewidth{0pt} 
\tablecaption{\label{catalog.tab}Catalog of \lya\ candidates.}
\tablehead{
\colhead{ID} &
\colhead{$\lambda$(\AA)\tablenotemark{a}} & 
\colhead{$z_{Ly\alpha}$} &
\colhead{R.A. (J2000)} & 
\colhead{Decl. (J2000)} &
\colhead{flux\tablenotemark{b}} &
\colhead{L$_{Ly\alpha}$\tablenotemark{c}} &
\colhead{class\tablenotemark{d}} &
\colhead{$z_{primary}$\tablenotemark{e}} 
}
\startdata
1112.73  &  7293 &  4.998 & 14:16:18.23 & +52:17:05.5 & 8.1$\times 10^{-18}$   &   2.1$\times 10^{42}$ & 1 & 0.7464 \\ 
1143.85  &  7505 &  5.1835& 14:15:20.55 & +52:07:38.6 & 7.8$\times 10^{-19}$   &   2.2$\times 10^{41}$ & 1 & 0.8789 \\ 
1150.116 &  7019 &  4.7694& 14:16:07.32 & +52:25:40.4 & 5.4$\times 10^{-18}$   &   1.3$\times 10^{42}$ & 3 & 0.7777 \\ 
2103.71  &  7029 &  4.7787& 16:47:01.01 & +34:49:56.7 & 7.8$\times 10^{-18}$   &   1.8$\times 10^{42}$ & 3 & 0.9727 \\ 
3307.61  &  8603 &  6.0559& 23:32:57.05 & +00:01:24.3 & 9.4$\times 10^{-18}$   &   3.8$\times 10^{42}$ & 2 & 0.8718 \\ 
4107.130 &  6972 &  4.7281& 02:27:28.03 & +00:35:34.3 & 9.1$\times 10^{-18}$   &   2.1$\times 10^{42}$ & 3 & 0.8311 \\ 
4110.123 &  6563 &  4.3975& 02:27:59.03 & +00:34:31.1 & 1.5$\times 10^{-17}$   &   3.0$\times 10^{42}$ & 2 & 0.9101 \\ 
4112.49  &  6668 &  4.4842& 02:27:58.09 & +00:26:17.3 & 1.2$\times 10^{-17}$   &   2.4$\times 10^{42}$ & 2 & 0.9071 \\ 
4147.18  &  6777 &  4.5750& 02:27:25.75 & +00:36:50.7 & 6.6$\times 10^{-18}$   &   1.4$\times 10^{42}$ & 2 & 0.9352 \\ 
4180.132 &  7682 &  5.3166& 02:26:08.48 & +00:42:32.0 & 5.5$\times 10^{-18}$   &   1.7$\times 10^{42}$ & 1 & 1.0393 \\ 
4218.94  &  6724 &  4.5252& 02:30:48.65 & +00:31:38.2 & 1.5$\times 10^{-17}$   &   3.2$\times 10^{42}$ & 3 & 0.8473 \\ 
4240.64  &  6769 &  4.5670& 02:28:38.04 & +00:41:07.4 & 4.9$\times 10^{-18}$   &   1.0$\times 10^{42}$ & 1 & 0.9691 \\ 
4243.121 &  6499 &  4.3452& 02:28:59.67 & +00:47:30.7 & 6.5$\times 10^{-18}$   &   1.2$\times 10^{42}$ & 1 & 0.9947 \\ 
4243.126 &  6996 &  4.7518& 02:29:01.06 & +00:48:17.2 & 7.2$\times 10^{-18}$   &   1.7$\times 10^{42}$ & 1 & 1.0213 \\ 
4256.30  &  6919 &  4.6883& 02:31:14.91 & +00:37:31.2 & 3.9$\times 10^{-18}$   &   8.8$\times 10^{41}$ & 1 & ---    \\ 
4257.9   &  7415 &  5.1026& 02:30:59.78 & +00:47:56.3 & 2.6$\times 10^{-18}$   &   7.0$\times 10^{41}$ & 1 & 0.9263 \\ 
4260.25  &  7908 &  5.5019& 02:31:14.92 & +00:37:36.1 & 4.9$\times 10^{-18}$   &   1.6$\times 10^{42}$ & 1 & 1.0370 \\ 
4280.73  &  6854 &  4.6359& 02:28:42.80 & +00:35:32.8 & 3.3$\times 10^{-18}$   &   7.3$\times 10^{41}$ & 3 & 0.9578 \\ 
4280.76  &  6865 &  4.6483& 02:28:40.70 & +00:35:56.8 & 1.1$\times 10^{-17}$   &   2.4$\times 10^{42}$ & 1 & 1.0045 \\ -

\enddata
\tablenotetext{a}{Wavelength of the peak of the emission line}
\tablenotetext{b}{In units of erg/s/cm$^2$ and not accounting for the (unknown) slit losses}
\tablenotetext{c}{In units of erg/s and not accounting for the (unknown) slit losses}
\tablenotetext{d}{Quality: 3=best: asymmetric line shape; 2=asymmetric line shape but object is very close to a foreground galaxy; 1=isolated emission line consistent with \lya, but without clear asymmetric line shape.}
\tablenotetext{e}{Redshift of the primary spectroscopic target}
\end{deluxetable}

\begin{deluxetable}{llc}
\tablewidth{0pt} 
\tablecaption{\label{coad-subsets.tab} LAE candidate confidence classes}
\tablehead{
\colhead{class} &
\colhead{description} & 
\colhead{number of objects} 
}
\startdata
3 & isolated object with asymmetric line profile characteristic of \lya; passes all the tests in \S~\ref{LyAsearch}; our best candidates 	& 5\\
2 & as class 1, but is close to foreground galaxy and so its flux could be inaccurate 							& 4\\
1 & line profile not clearly asymmetric, but otherwise object passes all the tests in \S~\ref{LyAsearch} 					& 10\\
\enddata
\end{deluxetable}





\clearpage

\appendix
\section{NOTES ON INDIVIDUAL OBJECTS}

All of the candidates discussed below have no associated spectral continuum either on the red or blue side unless otherwise specified.

{\it {Object 1112.73}}. This is a poor candidate due to the fact that the candidate's emission line is not asymmetric; its location in the slit is very close to the primary target galaxy.  The candidate, however, has no other associated lines. The object may have a faint counterpart in the B,R, \& I images which could rule out \lya\ as line ID.  We retain the object as a class 1 candidate.

{\it{Object 1143.85}} has a very weak emission line at 7505\,\AA\ which shows little to no asymmetry in the 2D and 1D spectra. The candidate is fairly well spatially separated from the primary target and is absent in the $BRI$ images. The faintness of the emission line gives the candidate less credibility as a LAE as it could possibly be  [\ion{O}{3}] at 5007\,\AA\ with undetected 4959\,\AA\ and 4861\,\AA.  We classify the object as a class 1 candidate.

{\it{Object 1150.116}} has a strong single emission line at 7020\,\AA\ that is moderately asymmetric in both the 1D and 2D spectra. The object is located well away from the primary target galaxy. There may be some continuum associated with the candidate on the redward side, especially a few hundred Angstroms redward of the potential \lya\ line. The candidate is undetected in the $BRI$ imaging. We classify it as a class 3 object.

{\it{Object 2103.71}} has a very strong emission line at 7030\,\AA\ that is strongly asymmetric in both the 1D and 2D spectra.  The object is located far from the primary DEEP2 target and is undetected in the $BRI$ images.  We classify it as a class 3 object.

{\it{Object 3307.61}} has an asymmetric emission line at 8603\,\AA\ in both the 1D and 2D spectra. Unfortunately, the object is very close to the primary target making it difficult to extract without contamination from the primary target continuum and thus  we cannot search for spectral continuum of this object.  The object is undetected in the $BRI$ images, but the location of the object is so close to the primary DEEP2 target that even a bright broadband counterpart would be difficult to detect.  We classify this object as a class 2 object. 

{\it{Object 4107.130}} has a strong emission line at 6973 \,\AA\ that is moderately asymmetric in both the 1D and 2D spectra. The candidate is not visible in $B$ or $R$, but could be present in the $I$ image at very low level of significance.  Class 3 object. 

{\it{Object 4110.123}} has a very strong emission line located at 6564\,\AA\ that is well separated from the primary target galaxy but is nearly superimposed on another serendipitous continuum source. This other serendipitous object has an  [\ion{O}{2}] emission line at 6843\,\AA\ and so is at $z$=0.836.  Our candidate 6564\,\AA\  line does not correspond to any known lines at $z$=0.836.  Moreover, the candidate Ly$\alpha$ line shows strong asymmetry in both the 1D and 2D spectra. The candidate appears absent in the $B$ image and is present in both $R$ and $I$ bands, although this is difficult to tell with certainty due to its proximity to the serendipitous [\ion{O}{2}] emitter. Flux measurement of this LAE candidate could potentially be affected by light from the [\ion{O}{2}] emitter and hence it is listed as class 2. 

{\it{Object 4112.49}} has a very strong emission line located at 6669\,\AA\ that is nearly superimposed on the primary DEEP2 target galaxy spectrum. The primary target is at a redshift of 0.8091 as determined from very strong  [\ion{O}{2}] emission at 6741\,\AA. The LAE candidate line is not matched by any known features at the redshift of the primary target.  The candidate Ly$\alpha$ line shows strong asymmetry in both the 1D and 2D spectra.  Broadband images are not useful in this case due to the proximity to the primary DEEP2 target. This is a class 2 candidate due to its proximity to the primary target. 

{\it{Object 4147.18}} has a strong emission line at 6778\,\AA\ that is strongly asymmetric in both the 1D and 2D spectra.  The candidate is also present in both the $R$ and $I$ images but absent in $B$-band, lending further credence to this candidate's bid as a LAE.   The potential Ly$\alpha$ line is nearly superimposed on another serendipitous spectrum which is an  [\ion{O}{2}] emitter at a redshift of 1.21; however the candidate LAE line does not match any lines at this redshift and hence we conclude that the object is not associated with the $z$=1.21 emitter.  While this is one of our best candidates based on its spectral shape, its apparent brightness may be affected by the foreground [\ion{O}{2}]-emitting galaxies and hence it is classified as a class 2 object. 

{\it{Object 4180.132}} has a very strong emission line located at 7683\,\AA\ that shows no asymmetry in either the 1D or 2D spectra, looking very much like a typical H$\beta$ or  [\ion{O}{3}] line. The position of the emission line is spatially  well separated from the primary target galaxy but does not appear to be present in any of the broadband images. Object classified as class 1. 

{\it{Object 4218.94}} has a very strong, very asymmetric line at 6724\,\AA\ that is well separated from the primary target. The emission line has a distinct asymmetric shape that is truncated on the blue end and extends in the red. The broadband images further lend credence to the supposition that this is a LAE, as the candidate appears in both the $R$ and $I$ bands but is absent in the $B$ band. This is quite possibly our best candidate.  Class 3. 

{\it{Object 4240.64}} contains a moderately bright emission line at 6769\,\AA\ that exhibits little to no asymmetry. This emission is fairly close to the primary DEEP2 target spatially, such that some of the target's continuum could have corrupted the spectral extraction of the LAE candidate. The candidate appears to be absent in all the broad band images.  Class 1 object. 

{\it{Object 4243.121}} has a very bright emission line at 6500\,\AA\ that shows little to no asymmetry. The broadband images show a galaxy at the position of this candidate that is present in $R$ and $I$ but absent in $B$.  However, it is difficult to tell whether this galaxy fell within the slit and thus whether it is associated with the line emitter in question.  Class 1 object. 

{\it{Object 4243.126}} has a moderately bright emission line at 6997\,\AA. The analysis of this candidate is difficult because a line of bad pixels runs directly through the emission line's center. However, it appears that the candidate's line shape has no asymmetry either in the 1D or 2D spectrum. The candidate is undetected in the $B$ band and may be present in the $R$ and $I$ bands.  Class 1 object. 

{\it{Object 4256.30}} has a bright emission line at 6919\,\AA\ that shows no asymmetry and looks similar to a normal H$\alpha$ or  [\ion{O}{3}] emission line. The candidate is quite close to the primary DEEP2 target spatially, making it challenging to tell whether the candidate is visible in the $BRI$ images. However, it does appear absent from these images.  Class 1 object.

{\it{Object 4257.9}} has a weak emission line at 7415\,\AA\ which has no asymmetry. The object does not appear to be present in any of the 3 broadband images. However, there may be an object in $R$ \& $I$ near the location of the candidate LAE and if the LAE candidate is associated with this galaxy then it would be ruled out as a LAE.  Class 1 object. 

{\it{Object 4260.25}} has a moderately bright emission line located at 7909\,\AA\ which has no asymmetry but --- given a non-detected continuum --- has a large equivalent width that is uncharacteristic of low-$z$ emission lines. The 1D spectrum is affected by bad night sky subtraction making it difficult to tell if there is any continuum associated with the emission line. The candidate is not present in either the $B$ or $R$ broadband images. There may be a galaxy present at the location of the candidate in the $I$ image, but this is not certain.  Class 1 object.

{\it{Object 4280.73}} has a strong emission line at 6855\AA\ that shows very well defined asymmetry in both the 1D and 2D spectra. The broadband images show no counterpart in $B$, $R$, or $I$.  Class 3 object. 

{\it{Object 4280.76}} has a strong emission line at 6866\AA\ which has little to no asymmetry but does have a large equivalent width making it unlikely that it is a line associated with a lower-$z$ galaxy. The candidate is absent from all three broadband images. Class 1 object.

\end{document}